\documentclass[apj]{emulateapj}
\usepackage{apjfonts}
\usepackage{amsbsy}

\newcommand{\msun}{M_{\odot}}
\newcommand{\kms}{{\rm km\, s}^{-1}}
\newcommand{\kpc}{{\rm kpc}}
\newcommand{\mpc}{{\rm Mpc}}
\newcommand{\hkpc}{{h^{-1}\kpc}}
\newcommand{\hmpc}{{h^{-1}\mpc}}
\newcommand{\hmsun}{{h^{-1}\msun}}
\newcommand{\up}[1]{{\rm #1}}

\begin{document}

\title{From Galaxy-Galaxy Lensing to Cosmological Parameters}

\author{Jaiyul Yoo\altaffilmark{1}, Jeremy L. Tinker\altaffilmark{2}, 
David H. Weinberg\altaffilmark{1}, Zheng Zheng\altaffilmark{3,4},
Neal Katz\altaffilmark{5}, \\ and Romeel Dav\'e\altaffilmark{6}}

\altaffiltext{1}{Department of Astronomy, The Ohio State University, 
140 West 18th Avenue, Columbus, OH 43210;
jaiyul@astronomy.ohio-state.edu, dhw@astronomy.ohio-state.edu}

\altaffiltext{2}{Kavli Institute for Cosmological Physics, 
University of Chicago, Chicago, IL 60637; tinker@cfcp.uchicago.edu}

\altaffiltext{3}{School of Natural Sciences, Institute for Advanced Study,
Einstein Drive, Princeton, NJ 08540; zhengz@ias.edu}

\altaffiltext{4}{Hubble Fellow}

\altaffiltext{5}{Departments of Physics and Astronomy, University of 
Massachusetts, Amherst, MA 01003; nsk@kaka.phast.umass.edu }

\altaffiltext{6}{Steward Observatory, University of Arizona, Tucson, AZ 85721;
rdave@as.arizona.edu}

\slugcomment{accepted for publication in The Astrophysical Journal}
\shorttitle{FROM GALAXY-GALAXY LENSING TO COSMOLOGICAL PARAMETERS}
\shortauthors{YOO ET AL.}

\begin{abstract}
Galaxy-galaxy lensing uses the weak distortion of background sources to
measure the mean excess surface density profile, $\Delta\Sigma(r)$,
around a sample of foreground lensing galaxies. We develop a method for
combining $\Delta\Sigma(r)$ with the galaxy-galaxy correlation
function $\xi_\up{gg}(r)$ to constrain the matter density parameter 
$\Omega_m$ and
the matter fluctuation amplitude $\sigma_8$, going beyond the widely used
linear biasing model to reach the level of accuracy demanded by current and
future measurements. We adopt the halo occupation distribution (HOD)
framework,
and we test its applicability to this problem by examining the effects of
replacing satellite galaxies in the halos of a smoothed particle
hydrodynamics (SPH) simulation with randomly selected dark matter particles
from the same halos. The difference
between dark matter and satellite galaxy radial profiles has a $\sim 10\%$
effect on $\Delta\Sigma(r)$ at $r<1~\hmpc$. However, if radial profiles are 
matched,
then the remaining impact of individual sub-halos around satellite galaxies
and environmental dependence of the HOD at fixed halo mass is $\lesssim 5\%$ in
$\Delta\Sigma(r)$ for $0.1 < r < 15~\hmpc$.  
We develop an analytic approximation
to $\Delta\Sigma(r)$ for a specified cosmological
model and galaxy HOD, improving on previous work with more accurate
treatments
of halo bias and halo exclusion. Tests against a suite of populated $N$-body
simulations show that the analytic approximation is accurate to a few
percent
or better over the range $0.1 < r < 20~\hmpc$. We use the analytic model to
investigate the dependence of $\Delta\Sigma(r)$ and 
the galaxy-matter correlation
function $\xi_\up{gm}(r)$ on 
$\Omega_m$ and $\sigma_8$, once HOD parameters for a
given cosmological model are pinned down by matching $\xi_\up{gg}(r)$. 
The linear
bias prediction that $\xi_\up{gm}(r)/\xi_\up{gg}(r)={\rm constant}$ 
is accurate for $r\gtrsim2~\hmpc$, but it fails at the $30-50\%$ 
level on smaller scales. The
scaling of $\Delta\Sigma(r)$ with cosmological
parameters, which we model as
$\Delta\Sigma(r) \propto \Omega_m^{\alpha(r)}\sigma_8^{\beta(r)}$, approaches
the linear
bias expectation $\alpha=\beta=1$ at $r\gtrsim10~\hmpc$, but $\alpha$ and
$\beta$
vary from 0.8 to 1.6 at smaller $r$. We calculate a fiducial 
$\Delta\Sigma(r)$ and scaling indices $\alpha(r)$ and $\beta(r)$ for galaxy 
samples that match
the observed number density and projected correlation function of Sloan
Digital
Sky Survey galaxies with $M_r \leq -20$ and $M_r \leq -21$. 
Galaxy-galaxy
lensing measurements for these samples can be combined with our predictions
to constrain $\Omega_m$ and $\sigma_8$, taking full advantage of the high
measurement precision on small and intermediate scales.
\end{abstract}


\section{Introduction}
\label{sec:int}
In the current paradigm of structure formation, galaxies form by the 
dissipative collapse of baryons in halos of cold dark matter (CDM).
Understanding the relation between the galaxy and dark matter distributions
is the key challenge in interpreting the observed clustering of galaxies. 
Large area imaging surveys have provided a new tool for 
untangling this relationship, galaxy-galaxy weak lensing, which 
uses the subtle distortion of background galaxy shapes to measure average mass 
profiles around samples of foreground galaxies.
The last few years have seen rapid growth in this field, with the
first tentative detections \citep{teresa} giving way to high signal-to-noise 
ratio measurements over a substantial
dynamic range (e.g., \citealt{fischer,tim,henk2,erin,mandel2}).

In a cosmological context, the strength of the galaxy-galaxy lensing signal 
for a given galaxy sample should depend mainly on the mean matter density 
$\Omega_m$ and the amplitude of dark matter fluctuations $\sigma_8$, since
increasing either parameter enhances the average amount of dark matter around 
galaxies and thereby amplifies the lensing signal.\footnote{Here $\sigma_8$
is the rms linear theory matter fluctuation in spheres of radius $8~\hmpc$,
with $h\equiv H_0/100~\kms\mpc^{-1}$.}
In this paper, we develop tools for constraining 
$\Omega_m$ and $\sigma_8$ with galaxy-galaxy lensing and galaxy clustering 
measurements, using halo occupation models of galaxy bias that are applicable 
from the linear regime into the fully non-linear regime. Our approach extends 
and complements earlier work by \citet{uros}, \citet{guzik1,guzik2}, 
\citet{iro}, and \citet{mandel}.

Galaxy-galaxy lensing measures the profiles of mean tangential shear around 
galaxies. With knowledge of source and lens redshift distributions, this
tangential shear can be converted to excess surface density,
\begin{equation}
\Delta\Sigma(r)\equiv{\bar\Sigma}(<r)-{\bar\Sigma}(r),
\label{eq:excess}
\end{equation}
where $\bar\Sigma(<r)$ is the mean surface density interior to the disk of 
projected radius $r$ and $\bar\Sigma(r)$ is the averaged surface
density in a thin annulus of the same radius \citep{jordi,erin}.\footnote{
We interchangeably use $r$ to refer to
a projected (two-dimensional) or a three-dimensional radius.}
The excess surface density profile is itself related to the galaxy-matter 
cross-correlation function $\xi_\up{gm}$ by
\begin{eqnarray}
\Delta\Sigma(r)&&=\rho_c\Omega_m\times \\
&&\left[{2\over r^2}\int_0^r\!\!\!
\int_{-\infty}^\infty\!\!\!r'\xi_\up{gm}\left(\sqrt{r'^2+z^2}\right)~dz~dr'
-\int_{-\infty}^\infty \!\!\!\xi_\up{gm}(r,z)dz\right], \nonumber
\label{eq:int}
\end{eqnarray}
where $\rho_c$ is the critical density of the universe.
\citet{johnston} discuss and test methods of inverting $\Delta\Sigma(r)$
to obtain the three-dimensional 
$\xi_\up{gm}(r)$. Here we treat $\Delta\Sigma(r)$
as the primary observable and concentrate on predicting it directly.

On large scales, where matter fluctuations are linear, the relation between 
the matter auto-correlation function $\xi_\up{mm}$, the galaxy-matter 
cross-correlation 
function $\xi_\up{gm}$, and the galaxy auto-correlation function 
$\xi_\up{gg}$ may be adequately described by the linear bias model,
\begin{equation}
\xi_\up{gg}=b^2\xi_\up{mm},
\label{eq:bias}
\end{equation}
\begin{equation}
\xi_\up{gm}=b\xi_\up{mm},
\label{eq:rbias}
\end{equation}
where the linear bias factor $b$ is the same in both equations \citep{nick}.
Thus, measurements of $\xi_\up{gg}$ and 
$\Delta\Sigma\propto\xi_\up{gm}\Omega_m$ can be combined
to yield $\Omega_m/b$. Since the amplitude of galaxy fluctuations is
$\sigma_\up{8,g}=b\sigma_8$ in the linear bias model,
this method in turn constrains the product $\sigma_8\Omega_m$. 
Redshift-space distortions of the galaxy power spectrum and the abundance of
rich galaxy clusters as a function of mass both depend on a parameter 
combination that is approximately $\sigma_8\Omega_m^{0.6}$ 
\citep{nick2,wef}, so the combination of 
galaxy-galaxy lensing with either of these measurements can break the 
degeneracy between $\sigma_8$ and $\Omega_m$. However, the linear bias 
approximation may break down on the scales $r\lesssim$ several $\hmpc$ where
$\Delta\Sigma(r)$ is measured with high precision. Moreover, if the relation
between galaxy and matter density contrasts is linear but stochastic, then the
linear bias factor $b$ in equation~(\ref{eq:rbias}) should be replaced by 
$br_\up{gm}$,
where $r_\up{gm}$ is the galaxy-matter cross-correlation coefficient 
\citep{ueli,dekel},
and the constrained combination becomes $\sigma_8\Omega_m r_\up{gm}$ 
even in the linear regime. The addition of 
$r_\up{gm}$ as a free parameter reduces the cosmological 
constraining power of the $\Delta\Sigma$ and $\xi_\up{gg}$ combination, and
restoring it requires an independent determination of $\xi_\up{mm}$. Cosmic
shear measurements can provide such a determination \citep{roger,jordi2,nick3},
but these measurements are challenging and suffer larger
systematic errors than galaxy-galaxy lensing.

\citet{henk3,henk2} used imaging and photometric redshift data from
the Red-Sequence Cluster Survey \citep{rcs} to measure aperture fluctuations
proportional to $\xi_\up{gg}$, $\xi_\up{gm}$, and $\xi_\up{mm}$. They provided
tentative evidence that $b$ and $r$ are each individually scale-dependent
but that the ratio $b/r_\up{gm}$ is approximately constant, with 
$b/r_\up{gm}\simeq1$ for $\Omega_m=0.3$.
Using the Sloan Digital Sky Survey (SDSS, \citealt{sloan}), which provides
spectroscopic redshifts of lens galaxies and image shapes and photometric
redshifts of source galaxies, 
\citet{erin} detected galaxy-galaxy lensing and measured the 
galaxy-matter correlation function from 0.025 to $10~\hmpc$
(see \citealt{fischer} and \citealt{tim} for earlier SDSS measurements,
and \citealt{mandel2} for more recent measurements at $r\leq2~\hmpc$).
They found that the galaxy-galaxy
correlation function and the galaxy-matter 
correlation function agree in shape, with an amplitude ratio that implies
$b/r_\up{gm}=(1.3\pm0.2)(\Omega_m/0.27)$ for galaxy samples of mean luminosity 
$\langle L\rangle\sim L_\star$.

To circumvent the limitations of the linear bias approximation, we model
galaxy-galaxy and galaxy-matter correlations using halo occupation methods,
following the lead of \citet{uros}, \citet{andreas}, and \citet{guzik2}.
The halo occupation distribution (HOD) provides a fully non-linear description
of the relation between galaxies and mass by specifying the probability 
$P(N|M)$ that a halo of virial mass $M$ contains $N$ galaxies of a particular
class, along with any spatial or velocity biases within individual 
halos.\footnote{Throughout this paper, we use the term ``halo'' to refer to a
dark matter structure of overdensity $\rho/\bar \rho_m\simeq200$, 
in approximate
dynamical equilibrium, which may contain a single bright galaxy or a group
or a cluster of galaxies.} Numerous authors have used this approach to compute
analytic approximations for galaxy and dark matter clustering statistics
(e.g., \citealt{chung,uros,roman,uros3,raby3,martin};
see review by \citealt{asantha}), and to model observed galaxy clustering
(e.g., \citealt{jing2,jing3,john2,chris,james2,mag,yang,por,idit2,idit4,
zheng2,kevork,col,goods,jeremy}).
Theoretical predictions for the HOD
of different galaxy types have been calculated using semi-analytic models,
hydrodynamic simulations, and high resolution $N$-body calculations that 
identify ``galaxies'' with dark matter substructures 
\citep{kau0,andrew1,whs,yosh,andreas2,andrey,andrew,zheng3}.

Our basic approach to modeling galaxy-galaxy lensing in the HOD framework is 
similar to that adopted by \citet{mtl,jeremy} for modeling mass-to-light
ratios and redshift-space distortions. Since measurements of the galaxy power 
spectrum, cosmic microwave background anisotropies, and the Ly$\alpha$ forest
yield tight constraints on the shape of the linear matter power spectrum
$P_\up{lin}(k)$ (see, e.g., \citealt{wmap2,max1,2dfn,uros4}), 
we take this shape
to be fixed and investigate the parameter space spanned
by $\Omega_m$ and $\sigma_8$. For a given choice of $\Omega_m$ and $\sigma_8$,
we first choose HOD parameters to match observations of the projected 
galaxy correlation function $w_p(r_p)$ (see, e.g., \citealt{idit2,idit4}).
We then predict the excess surface density profile
$\Delta\Sigma(r)$ for this combination of $\Omega_m$, $\sigma_8$, and HOD.
Comparison to galaxy-galaxy lensing measurements then determines the acceptable
combinations of $\Omega_m$ and $\sigma_8$. We impose the observed galaxy-galaxy
correlation function as a constraint on the HOD, instead of taking ratios as in
the linear bias analysis. There is no need for an unknown cross-correlation
coefficient $r_\up{gm}$ 
because any ``stochasticity'' between galaxy and mass density
fields is automatically incorporated in the HOD calculation.
Our strategy complements that of \citet{guzik1,guzik2} and \citet{mandel},
who focus on constraining halo masses, halo profiles, and satellite 
fractions using $\Delta\Sigma(r)$ alone, rather than constraining $\Omega_m$
and $\sigma_8$ from the combination of $\Delta\Sigma(r)$ and $\xi_\up{gg}(r)$.

Our eventual conclusions about the cosmological constraining power of 
galaxy-galaxy lensing measurements rest on an analytic model for computing
$\Delta\Sigma(r)$ given $P_\up{lin}(k)$, $\Omega_m$, $\sigma_8$, and the galaxy
HOD. This model is similar in spirit to that of \citet{guzik2}, but it differs
in many details, in part because we define the calculational problem in 
different terms.
We test the analytic model against numerical calculations, in which
we use a specified HOD to populate the halos of $N$-body simulations, 
placing ``central'' galaxies at the halo potential minimum and ``satellite''
galaxies at the locations of randomly selected dark matter particles. Both
our analytic model and our method of populating $N$-body halos ignore the
impact of dark matter $subhalos$ around the individual
satellite galaxies orbiting in a larger halo. To begin, therefore, we test the
validity of the ``populated halo'' approach itself, by comparing 
$\Delta\Sigma(r)$ for the galaxy population of a smoothed particle 
hydrodynamics
(SPH) simulation to that found by populating the dark matter halos of this
simulation with ``galaxies'' placed on randomly selected dark matter particles.
This test shows that satellite subhalos have minimal impact on
$\Delta\Sigma(r)$ and that the populated halo approach is acceptable for our 
purpose. We also show that any environmental dependence of halo galaxy content
at fixed halo mass \citep{gao,harker} has little discernible impact on the 
galaxy-galaxy or galaxy-matter correlation functions in our SPH simulations.
More generally, our SPH and $N$-body tests indicate that the analytic 
model should be accurate at the $5-10\%$ level on scales $r\gtrsim0.1~\hmpc$. 
This level of accuracy is acceptable for present purposes, since the current 
measurement errors are typically $\gtrsim25\%$ per radial bin
(e.g., \citealt{erin}), but still higher accuracy will be needed in the long
term.

In our $N$-body and analytical calculations, we use HOD parameters for SDSS 
galaxy samples with absolute-magnitude limits $M_r\leq-20$ and $M_r\leq-21$ 
\citep{idit3} for purposes of illustration.\footnote{Throughout the paper,
we quote absolute magnitudes for $h=1$; more generally, these thresholds
correspond to $M_r-5\log h$.} 
The results presented in 
\S~\ref{sec:dep} therefore yield predictions of the weak lensing signal for 
these galaxy samples as a function of $\Omega_m$ and $\sigma_8$. The analytic 
model can be used to make predictions for other galaxy samples given 
measurements of the projected correlation function as input.

\begin{figure*}
\centerline{\epsfxsize=6.5truein\epsffile{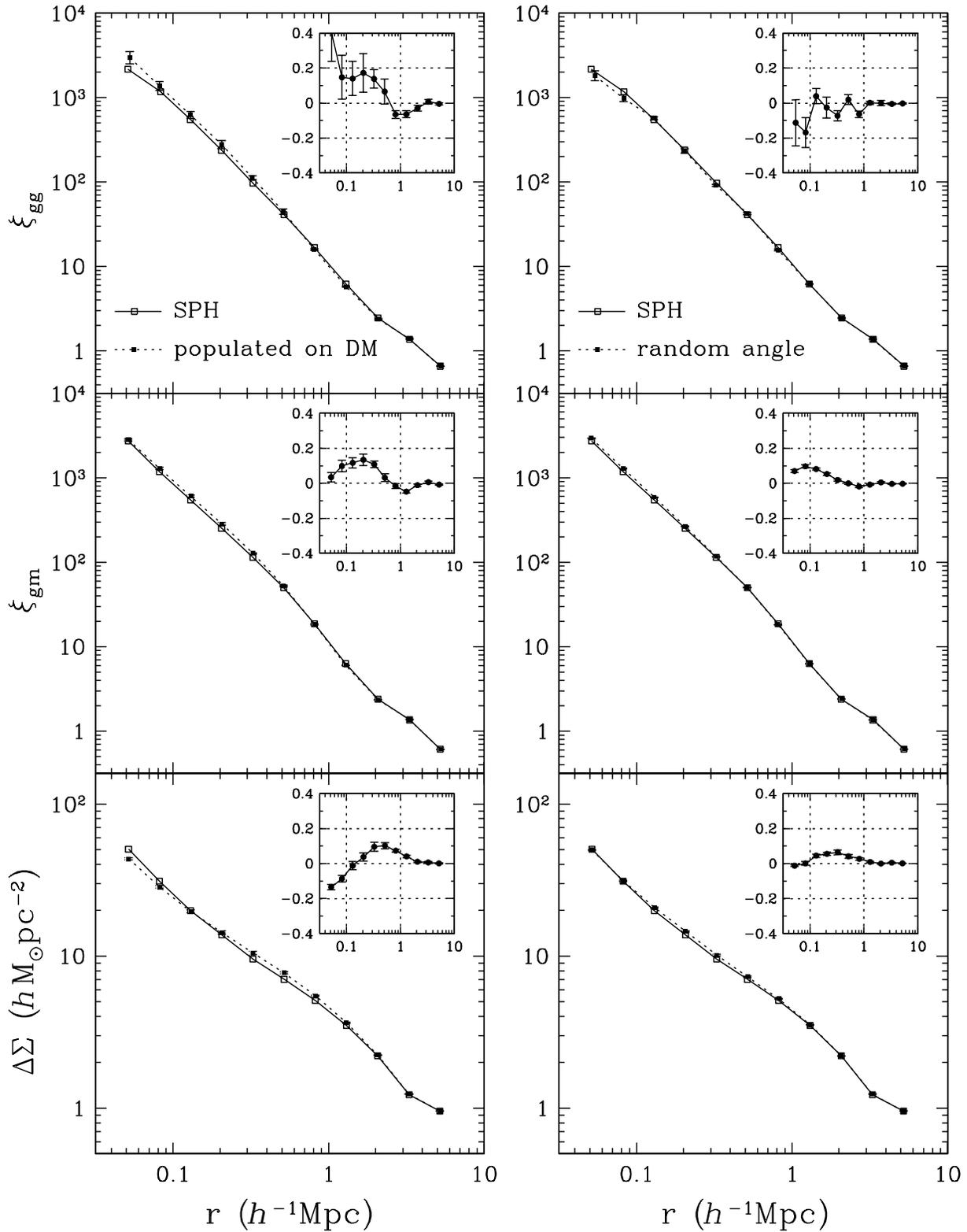}}
\caption{Galaxy-galaxy correlation functions ({\it top panels}), galaxy-matter
correlation functions ({\it middle panels}), and $\Delta\Sigma(r)$ profiles
({\it bottom panels}) for the true galaxy population of an SPH simulation
({\it solid lines}) and the populated dark matter halos of this simulation
({\it dotted lines}, see text). Inset panels show the fractional difference
between the SPH and populated halo results. In the left-hand panels, satellite
galaxies in the populated halos are placed on randomly selected dark matter
particles, while in the right-hand panels they are forced to follow the
radial profile of satellite galaxies in the SPH simulation.}
\label{fig:sph}
\end{figure*}

\begin{figure}[t]
\centerline{\epsfxsize=3.5truein\epsffile{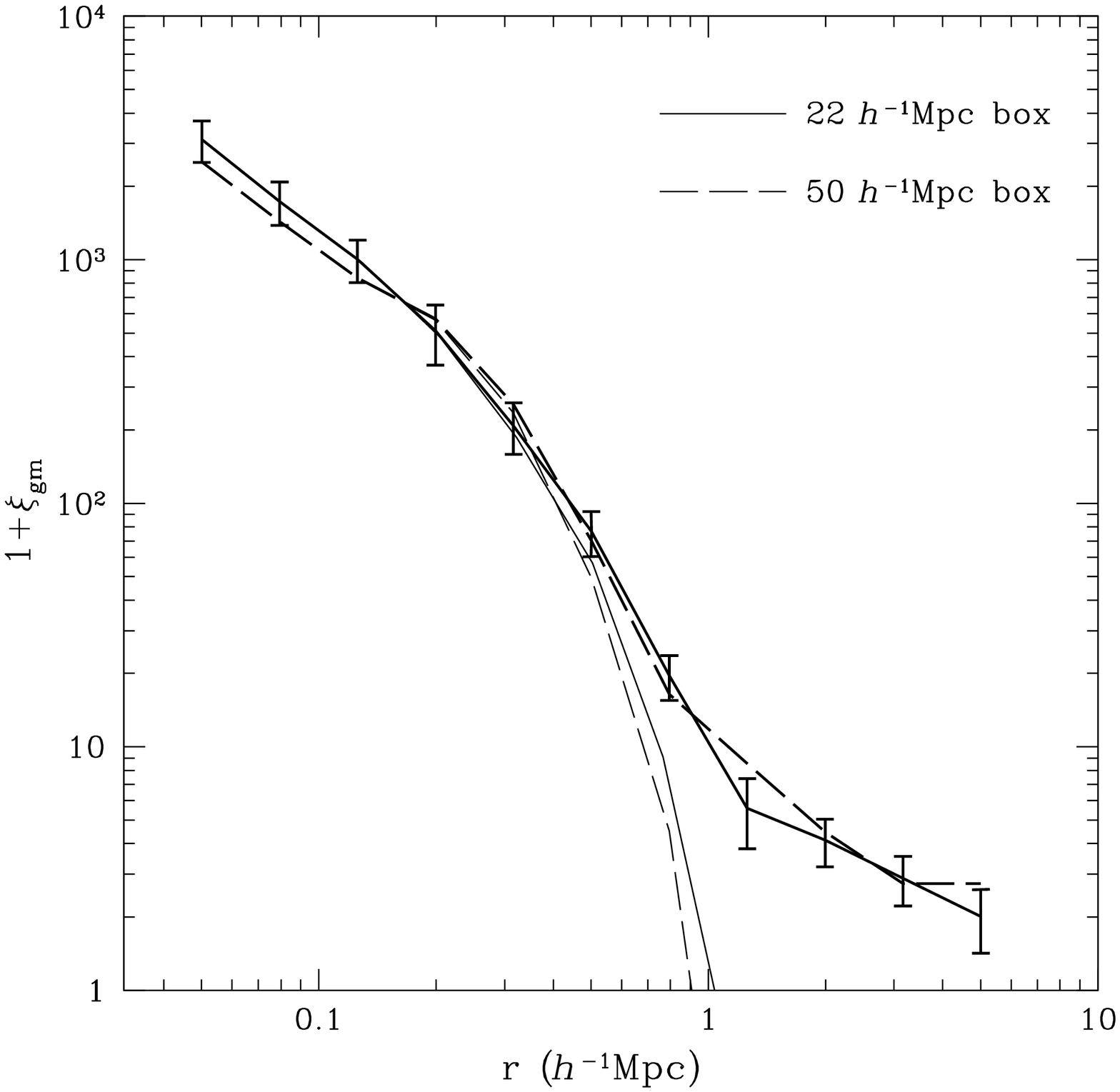}}
\caption{Comparison of galaxy-matter correlation functions for satellite 
galaxies of our standard SPH simulations ({\it dashed lines}), which uses
144$^3$ particles in a $50~\hmpc$ box, to those of a higher resolution, smaller
volume simulation (128$^3$ particles in a $22.222~\hmpc$ box, 
{\it solid lines}). In both simulations we select halos in the mass range
$6\times10^{12}\hmsun\lesssim M\lesssim2\times10^{13}\hmsun$ and satellite
galaxies above the $3.5\times10^{10}~\hmsun$ resolution limit of the larger
volume run. Heavy lines show the full satellite-matter cross-correlation 
function, while light lines include only matter in the satellite's parent
halo (the one-halo term). Error bars are computed for the smaller simulation
via bootstrap resampling.}
\label{fig:sub}
\end{figure}

\section{SPH Galaxies versus Populated Halos}
\label{sec:comp}
To test the validity of our $N$-body method for calculating galaxy-galaxy
lensing predictions (see \S~\ref{sec:numerical}), we first examine an SPH
simulation of a $\Lambda$CDM (inflationary cold dark matter with a cosmological
constant) universe. This simulation is described in detail by \citet{dhw},
who, among other things, present predicted galaxy-matter correlations and 
compare them to recent observations. Here we want to know whether the 
individual dark matter subhalos retained by baryonic galaxies in groups and
clusters make an important contribution to the galaxy-galaxy lensing signal.

In brief, the simulation uses Parallel TreeSPH \citep{lars,kwh,dave} to model
a $50~\hmpc$ comoving cube with $144^3$ dark matter particles and $144^3$ gas
particles. The cosmological parameters are $\Omega_m=0.4$, 
$\Omega_\Lambda=0.6$, $h=0.65$, $n=0.95$, $\Omega_b{\it h}^2=0.02$, and
$\sigma_8=0.80$. The gravitational forces are softened with a $10~\hkpc$ 
(comoving) spline kernel. Radiative cooling leads to the formation of dense
baryonic clumps \citep{khw,evr}, which form stars according to the algorithm
described by \citet{kwh}. Galaxies are identified by applying the SKID
(Spline-Kernel Interpolated DENMAX; see \citealt{kwh}) algorithm to the 
population of stars and cold, dense gas particles.\footnote{We use the 
implementation of SKID by J. Stadel \& T. Quinn, available at \\
{\tt http://www-hpcc.astro.washington.edu/tools/skid.html}}
Tests with simulations of varying resolution show that the simulated galaxy
population is complete above a baryonic mass (stars plus cold, dense gas)
of $\sim$64 $m_\up{SPH}$, corresponding to $5.4\times10^{10}\msun$ 
($3.5\times10^{10}~\hmsun$) for this simulation. The space density of galaxies
above this mass threshold is ${\bar n}_g=0.02~(\hmpc)^{-3}$, corresponding to
that of observed galaxies with $M_r\leq-18.6$ 
($L>0.18~L_\star$; \citealt{lstar}).
We use this mass-thresholded galaxy sample for the tests below.

We identify dark matter halos by applying the friends-of-friends algorithm
(FOF; \citealt{fof}) to the dark matter particle distribution, with a linking
length of 0.2 times the mean interparticle separation, or $70~\hkpc$.
We associate each SPH galaxy with the halo containing the dark matter particle
closest to its center of mass. To create ``populated halo'' galaxy catalogs,
we replace the SPH galaxies in each halo with an equal number of artificial
galaxies positioned on dark matter particles. The first ``central'' galaxy of
each occupied halo is placed at the location of 
the dark matter particle with
lowest potential energy (computed using only halo members).
Any additional, ``satellite'' galaxies
are placed on randomly selected dark matter particles. Satellites therefore
follow the radial profile of dark matter within each halo, while any detailed
association between satellites and the centers of dark matter subhalos is
erased.

The left panels of Figure~\ref{fig:sph} compare the galaxy-galaxy correlation
functions, galaxy-matter correlation functions, and excess surface density
profiles of the SPH galaxies and the populated halo galaxy catalogs. Results
for the populated halos are an average over 10 realizations of the galaxy
locations, and error bars show the dispersion among the ten realizations.
(Note that these do {\it not} 
represent the uncertainty on the mean, which would be
a factor of three smaller, and they do not include the uncertainty 
owing to the
finite simulation volume, since we are comparing galaxy catalogs in the same
volume). The galaxy-galaxy and galaxy-matter correlations of the two catalogs
are very similar at $r\gtrsim0.5~\hmpc$, while at smaller separations the
populated halo catalog has correlations that are stronger by up to 20\%. The
excess surface density profile is calculated by directly counting galaxy-dark
matter particle pairs in projection to compute $\bar\Sigma(<r)$ and
$\Sigma(r)$, not by integrating the three-dimensional $\xi_\up{gm}(r)$.
We count all projected pairs through the $50~\hmpc$ box and average
results from the three orthogonal projections of the box; noise from
uncorrelated foreground and background particles cancels out because we have
many galaxy targets.
Relative to the SPH galaxy catalog, $\Delta\Sigma(r)$ for the populated halo
catalog starts about 10\% low, rises to 10\% above at $r\sim0.5~\hmpc$, then
agrees closely beyond $r\sim1~\hmpc$.

The modest deviations between the SPH galaxy and populated halo results could 
reflect either the impact of satellite subhalos or differences between
the radial profiles of SPH satellites and dark matter. To separate the two
effects, we adopt a different method of populating halos that ensures identical
radial profiles, by placing each 
satellite at the radial distance of the corresponding SPH satellite but 
choosing a random orientation for the radius vector. Results are shown on the
right panels of Figure~\ref{fig:sph}. The differences in $\xi_\up{gg}(r)$,
$\xi_\up{gm}(r)$, and $\Delta\Sigma(r)$ are greatly reduced, demonstrating that
they arise mainly from the different radial profiles of SPH satellite galaxies
and dark matter;
specifically, the SPH satellites are less concentrated towards the halo
center than the dark matter. With matched radial profiles,
the populated halos still have slightly larger 
$\xi_\up{gm}(r)$ at $r\lesssim0.1~\hmpc$, in part because there are usually
offsets of this magnitude between the location of the SPH central galaxy and 
the position of the most bound dark matter particle. However, the differences 
in $\Delta\Sigma(r)$ are now smaller than 10\% at all $r$.

We conclude that it is safe to ignore the subhalos of individual satellite
galaxies when computing $\Delta\Sigma(r)$ for a full galaxy sample. Indeed,
the remaining residuals in Figure~\ref{fig:sph}, a consequence of the central
galaxy offsets mentioned above, are opposite in sign to those expected from
satellite subhalos. Satellites in the SPH simulation $do$ reside in individual
dark matter subhalos \citep{dhw2},
but these are tidally truncated, and at small separations
$\Delta\Sigma(r)$ is dominated by the contribution of the more numerous,
central galaxies (see \S~\ref{sec:dissection} below). The small impact of 
satellite subhalos on $\Delta\Sigma(r)$ is therefore unsurprising, and was
anticipated by earlier analytic modeling \citep{guzik2,mandel}.

Nonetheless, one might worry that the absence of $any$ subhalo signal in 
Figure~\ref{fig:sph} is an artifact of our simulation's finite mass resolution,
leading to an artificially high degree of tidal truncation. To test this
possibility, we compare results from this simulation to those of a simulation 
of the same cosmological model with a factor of eight higher mass resolution
but smaller volume. This simulation uses 128$^3$ dark matter particles and
128$^3$ gas particles in a volume $22.222~\hmpc$ on a side. In each simulation,
we select all halos in the mass range 
$6\times10^{12}~\hmsun\lesssim M\lesssim 2\times10^{13}~\hmsun$ and 
measure the
galaxy-matter correlation function for $satellites$ in these halos above the
baryonic mass threshold of the larger volume, lower resolution simulation.
The $50~\hmpc$ box contains 78 halos and 188 satellite galaxies satisfying
these cuts, while the $22.222~\hmpc$ box contains 12 halos and 33 satellites. 
As shown in Figure~\ref{fig:sub}, the satellite galaxy-matter
correlations are equal in the two simulations to within the statistical errors,
which are estimated by bootstrap resampling of the galaxies in the smaller
simulation. The average mass profiles around satellites are therefore robust
over a factor of eight in mass resolution.

Standard HOD calculations assume that the halo occupation function $P(N|M)$
has no direct dependence on a halo's larger scale environment. This assumption
is motivated by the excursion set derivation of the Extended Press-Schechter
formalism \citep{bcek}, which,
in its simplest form,
predicts that a halo's formation history is
uncorrelated with its environment at fixed mass \citep{les}. 
The correlation of
galaxy properties with large scale environment emerges indirectly from the 
correlation with halo mass because high mass halos are more common in dense
environments. \citet{blanton}
showed that the observed correlation of red galaxy fraction
with overdensity measured at $6~\hmpc$ is entirely accounted for by the 
correlation with overdensity measured at the $1~\hmpc$ scale characteristic of
individual large halos. However, while early $N$-body studies showed at most
weak correlations between halo formation time and environment at fixed mass
for halos with $M\gtrsim10^{13}\hmsun$ \citep{lemson99,sheth04}, \citet{gao}
have recently shown that
there is a much stronger correlation for lower mass halos, with the older halos
being more strongly clustered (see also 
 \citealt{st02,harker}
discuss the potential origin of environmental correlations in the excursion 
set formalism). 
\citet{andreas2},
examining the same SPH simulation and 
galaxy sample that we have used here, showed that the mean number of galaxies 
as a function of halo mass, $\langle N\rangle_M$, is independent of halo 
environment within the statistical uncertainties imposed by the finite 
simulation volume. However, in light of Gao et al.'s (2005) result, we have
carried out an experiment to explicitly examine the possible impact of 
environmental dependence of $P(N|M)$ on galaxy-galaxy and galaxy-matter
correlations.

In the populated halo calculations shown in Figure~\ref{fig:sph}, the number
of galaxies assigned to each halo is equal to the number of SPH galaxies,
so any environmental dependence predicted by the SPH simulation is also built
into the populated halo distribution. We eliminate the environmental dependence
by shuffling the galaxy populations among halos of similar mass. Specifically,
we sort the halos by mass and replace the number of galaxies $N_i$ in the halo
of rank $i$ with the number $N_{i+1}$
in halo $i+1$, then recalculate $\xi_\up{gg}(r)$, $\xi_\up{gm}(r)$, and 
$\Delta\Sigma(r)$, averaging over ten realizations of galaxy positions 
$within$ halos. We repeat the exercise with the substitutions
$N_{i+1}\rightarrow N_{i+2}$, $N_{i-1}$, and $N_{i-2}$ so that we can average 
over four different halo shufflings and compute the statistical error on the 
mean. The sampling of the halo mass function is sparse for the highest mass 
halos in the simulation, so at high masses we cannot exchange the galaxy 
contents
of halos without significantly changing $P(N|M)$ itself. We therefore keep the
galaxy populations of the $N_\up{fix}$ most massive 
halos fixed, with $N_\up{fix}=5$
or 20. For $N_\up{fix}=5$, we are shuffling the contents of all halos with
$M<9.0\times10^{13}~\hmsun$, and for $N_\up{fix}=20$ we are shuffling the 
contents of all halos with $M<4.6\times10^{13}~\hmsun$.

\begin{figure}[t]
\centerline{\epsfxsize=3.6truein\epsffile{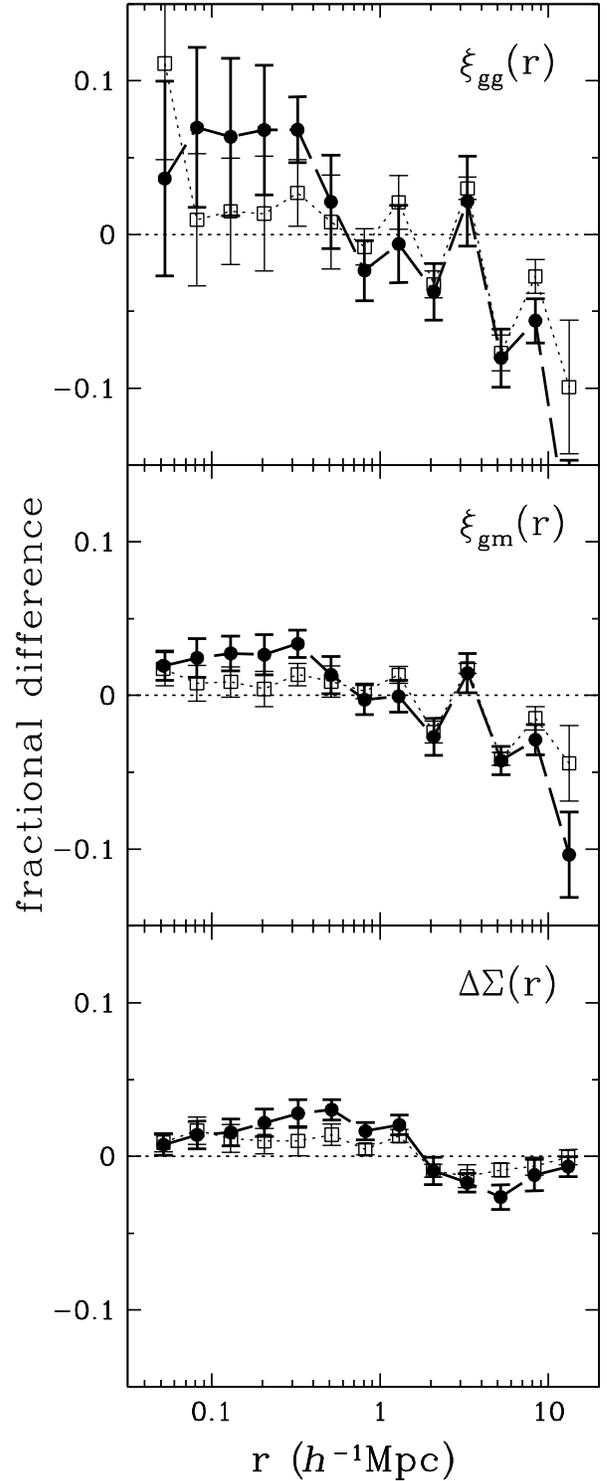}}
\caption{Possible impact of environmental variation of the HOD on
$\xi_\up{gg}(r)$, $\xi_\up{gm}(r)$, and $\Delta\Sigma(r)$. We shuffle the 
occupation numbers of halos of similar mass, leaving the populations of the
five ({\it filled circles}) or 20 ({\it open squares}) most massive halos
unchanged. Plots show the fractional difference between the shuffled halo 
results and the original results. Error bars in the points show the uncertainty
in the mean calculated from four different shufflings (see text).}
\label{fig:gao}
\end{figure}

Figure~\ref{fig:gao} plots the fractional difference in $\xi_\up{gg}(r)$,
$\xi_\up{gm}(r)$, and $\Delta\Sigma(r)$ between the shuffled halo realizations
and the original populated halos. We use the populated halos as the comparison
standard rather than the SPH galaxies so that we can isolate the impact of 
environmental dependence of $P(N|M)$. Error bars show the error on the mean 
from the four shufflings, but recall that we have only one realization of the
original populated halos. For $N_\up{fix}=5$, there is a 5\% increase on
$\xi_\up{gg}(r)$ at $r\lesssim0.5~\hmpc$.  However, these scales lie in the 
1-halo regime where environmental variation of $P(N|M)$ should have no
impact at all, so the increase is probably a statistical fluctuation
that reflects the particular sizes and concentrations of the halos present
in the simulation. It is only
slightly larger than the 1$\sigma$ error bars, and the errors from point to
point are highly correlated. For $N_\up{fix}=20$, the changes in 
$\xi_\up{gg}(r)$ are less than 3\% over the range 
$0.08~\hmpc\lesssim r\lesssim3~\hmpc$. The three points at $r\gtrsim5~\hmpc$
are depressed by $\sim5\%$ on average, which suggests that shuffling may 
slightly lower the large scale galaxy bias factor, but the statistical 
significance of this depression is difficult to assess with a single $50~\hmpc$
simulation.

Shuffling changes $\xi_\up{gm}(r)$ by less than 5\%, usually much less, except
for the largest scale point with $N_\up{fix}=5$. Most significantly for our
present purposes, the changes in $\Delta\Sigma(r)$ are at most $\sim2\%$ for
$N_\up{fix}=20$ at all scales, and only slightly larger for $N_\up{fix}=5$.
We conclude that ignoring any possible environmental dependence of $P(N|M)$ has
minimal impact on the calculation of galaxy-galaxy lensing observables for a
given cosmology and HOD, for a galaxy sample defined by a threshold in
baryonic mass. There
could be a few percent effect on the large scale bias of the galaxy-galaxy
correlation function, which might lead to small errors in inferring the HOD
from observations of $\xi_\up{gg}(r)$. Assessing the importance of this effect
will require larger simulations. \citet{cro2} have carried
out a similar shuffling experiment for semi-analytic galaxy populations in the
$500~\hmpc$ Millennium Run simulation \citep{croton}, and they find few percent
changes in large scale bias for galaxy samples defined by 
 thresholds in mass or absolute magnitude
(it was the hearing about
their shuffling experiment that inspired us to carry out our own).

\begin{deluxetable}{ccccccc}
\tablewidth{0pt}
\tablecaption{Parameters of {\scriptsize GADGET} 
Simulations and HOD Parameters}
\tablehead{\colhead{Model} & \colhead{$\Omega_m$} & \colhead{$\sigma_8$}
& \colhead{$M_\up{min}(\hmsun)$} & \colhead{$M_1(\hmsun)$}
& \colhead{$\alpha_\up{sat}$} & \colhead{$\Delta_\up{vir}$}}
\startdata
1 & 0.10 & 0.95 & $3.81\times10^{11}$ & $1.22\times10^{12}$ & 1.04 & 240\\
2 & 0.16 & 0.90 & $6.27\times10^{11}$ & $1.73\times10^{13}$ & 1.03 & 210\\
3 & 0.30 & 0.80 & $1.17\times10^{12}$ & $2.65\times10^{13}$ & 1.04 & 190\\
4 & 0.47 & 0.69 & $1.87\times10^{12}$ & $3.49\times10^{13}$ & 1.05 & 165\\
5 & 0.63 & 0.60 & $2.41\times10^{12}$ & $3.85\times10^{13}$ & 1.09 & 155\\
\enddata
\tablecomments{The HOD parameters of the
fiducial model ({\it Model~3}) are chosen to reproduce the
same clustering of the SDSS galaxy sample of $M_r\leq-20$ and to match the 
number 
density of galaxies $\bar n_g=5.74\times10^{-3}~(\hmpc)^{-3}$. 
HOD parameters of the other
models are scaled with $\Omega_m$ from the HOD parameters of the 
fiducial model and adjusted to match $\xi_\up{gg}$ and $\bar n_g$.}
\label{tab:par}
\end{deluxetable}

\begin{figure*}[t]
\centerline{\epsfxsize=5.5truein\epsffile{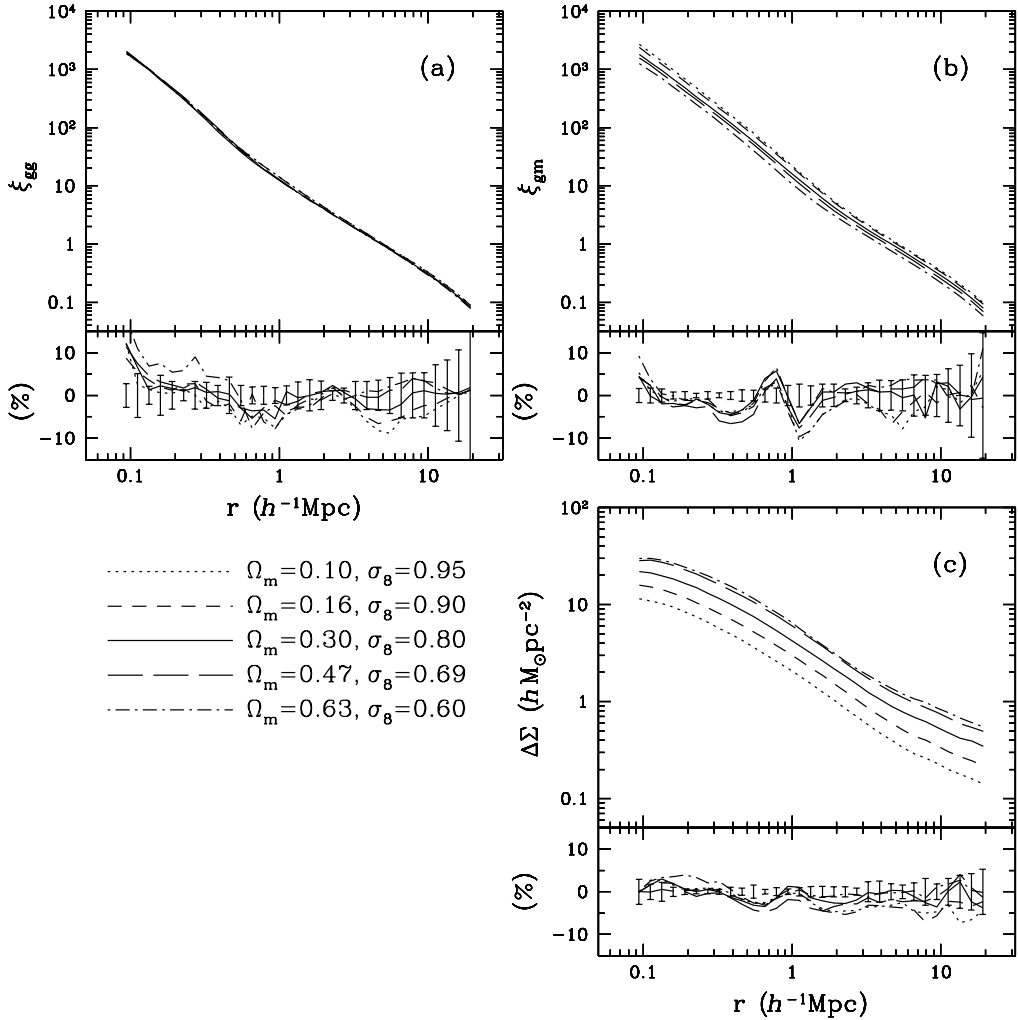}}
\caption{Large panels show $N$-body results for $\xi_\up{gg}(r)$, 
$\xi_\up{gm}(r)$, and $\Delta\Sigma(r)$ for the five cosmological parameter
combinations indicated in the legend and detailed in Table~\ref{tab:par}.
Attached bottom panels show the fractional difference between the analytic
model calculations and the simulation results. Error bars represent
fractional statistical uncertainty on the $N$-body results for the central 
model ($\Omega_m=0.3$ and $\sigma_8=0.8$, $solid~lines$), computed from the 
error on the mean of the five simulations.}
\label{fig:comp}
\end{figure*}

\section{$N$-Body Simulations}
\label{sec:numerical}
To help us develop and test our analytic model, we have carried out five
$N$-body simulations of a $\Lambda$CDM universe using {\scriptsize GADGET}
\citep{gadget}. Each simulation begins at expansion factor $a=0.01$ with
a scale-invariant ($n=1$) fluctuation spectrum modulated by the transfer 
function of \citet{george} with shape parameter $\Gamma=0.2$. 
 Our analytic model calculations in \S~\ref{sec:dep} use the
{\scriptsize CMBFAST} transfer function \citep{cmbfast}, 
which represents cosmological predictions more accurately, but the 
\citet{george} representation should be adequate for calibrating and testing
the analytic model itself.
The simulations
end at $a=1.0$, when $\Omega_m=0.1$, $\Omega_\Lambda=0.9$, and the linear 
theory normalization of the power spectrum is $\sigma_8=0.95$. We use earlier
outputs from the same simulations to represent models with the cosmological
parameter combinations listed in Table~\ref{tab:par}:
($\Omega_m$, $\sigma_8$)=(0.16, 0.90), (0.30, 0.80), (0.48, 0.69), and
(0.63, 0.60). Since we are adopting a fixed, observationally motivated form of
the power spectrum instead of changing its shape with $\Omega_m$, this 
procedure is exact. We would obtain the same results if we ran a separate
simulation for each model but started it at expansion factor 
$a=0.01/a_\up{out}$, where $a_\up{out}=$~0.84, 0.64, 0.49, 0.40 for the four
$(\Omega_m,~\sigma_8)$ combinations.
We refer the reader to \citet{mtl,jeremy} for the simulation details.

Our simulations use 360$^3$ particles to model a volume $253~\hmpc$ (comoving)
on a side. The dark matter particle mass is $9.6\times10^{10}\Omega_m\hmsun$.
We choose the mass resolution so that the lowest mass halos that 
host galaxies with $M_r\leq-20$, according to our HOD fits (see below),
contain at least
32 particles. The gravitational force resolution is $\epsilon=70~\hkpc$
(this is the approximate Plummer-equivalent value).
The five simulations are identical except for the random number seed used to
generate the initial conditions. We identify dark matter halos using FOF with
a linking length equal to 0.2 times the mean interparticle separation, or 
$140~\hkpc$,
and set the halo mass equal to the total mass of the linked particles.

We populate the $N$-body halos with galaxies using HOD parameters that are
designed to reproduce the mean space density and projected correlation function
of SDSS galaxies with $M_r\leq-20$, as measured by \citet{idit3}.
The adopted form of the HOD is motivated by the results of \citet{andrey}
and \citet{zheng3}. Halos below some minimum
mass $M_\up{min}$ are devoid of galaxies.
All halos above $M_\up{min}$ have a central
galaxy, which is placed at the position of the dark matter particle with the
lowest potential energy in each halo.
The number of satellite galaxies is drawn from a 
Poisson distribution with mean $(M/M_1)^{\alpha_\up{sat}}$. 
Each satellite galaxy is 
placed on a randomly selected dark matter particle from the halo.
Table~\ref{tab:par} lists the values of $M_\up{min}$, $M_1$, and 
$\alpha_\up{sat}$ for
our five ($\Omega_m$, $\sigma_8$) combinations. Further details of the fitting
procedure are given by \citet{jeremy}.
The specifics of the parametrization and details 
of the fitting method are not important to our purposes here, since we will
test the analytic model predictions using the same HOD parameters applied to
the simulations. However, these parameter choices ensure a galaxy population
with realistic clustering properties.

Figure~\ref{fig:comp} shows $\xi_\up{gg}(r)$, $\xi_\up{gm}(r)$, and 
$\Delta\Sigma(r)$ for the five $N$-body models. The five galaxy-galaxy 
correlation functions are nearly identical by construction, though with the
HOD parameters at our disposal it is not possible to exactly match the observed
correlation function over our full range of $\sigma_8$. The galaxy-matter 
correlation function is higher for the more strongly clustered, high $\sigma_8$
models, as expected. However, since $\Delta\Sigma(r)$ scales (approximately)
with $\Omega_m\sigma_8$, and $\Omega_m$ falls faster than $\sigma_8$ grows
in our simulation outputs, the order of models is reversed on the 
$\Delta\Sigma(r)$ panel. We discuss the comparison between the $N$-body and 
analytic model results in the following section.

\section{Analytic Modeling of Galaxy-Matter Clustering}
\label{sec:analytic}

\subsection{Formulation and Tests}
\label{sec:test}
Our analytic method of calculating $\xi_\up{gm}(r)$ for a given cosmology
and HOD is based on the methods that \citet{zheng2} and 
Tinker~et~al. (2005a, see Appendix~B) used to calculate the galaxy-galaxy
correlation function. These methods are based, in turn, on ideas introduced
by \citet{bob}, \citet{chung}, \citet{uros}, \citet{john2} and \citet{roman}. 
We present a full technical
description of our $\xi_\up{gm}$ calculation here but refer the reader to
these earlier works for more general discussion. Our galaxy-galaxy correlation
calculations follow \citet{mtl}, with ellipsoidal halo exclusion.

Contributions to $\xi_\up{gm}$ can come from galaxy-matter 
pairs\footnote{By which we mean pairs of galaxies and dark matter particles.}
residing in
a single halo or in two distinct halos. We separate these two contributions as
\begin{equation}
1+\xi_\up{gm}(r)=\left[1+\xi^\up{1h}_\up{gm}
(r)\right]+\left[1+\xi^\up{2h}_\up{gm}(r)
\right],
\end{equation}
noting that it is pair counts (proportional to $1+\xi_\up{gm}$) that add rather
than the correlations $\xi_\up{gm}$ themselves. The one-halo contribution is
\begin{equation}
1+\xi_\up{gm}^\up{1h}(r)={1\over4\pi r^2{\bar n}_g}\int_{M_\up{min}}
^\infty\!\!\!\!\!
dM{dn\over dM}\langle N\rangle_M{M\over{\bar\rho_m}}{1\over2R_\up{vir}}
\up{F}'\left({r\over2R_\up{vir}}\right),
\label{eq:1h}
\end{equation}
where $dn/dM$ is the halo mass function \citep{ps,raby,jenkins}, 
$\langle N\rangle_M$ is the mean number of galaxies in halos of mass $M$,
$\bar\rho_m$ is
the mean mass density, and $F(r/2R_\up{vir})$ is the average fraction of 
galaxy-matter pairs in halos of mass $M$ and virial radius $R_\up{vir}$ that
have separation less than $r$ (\citealt{andreas}; $F'(x)$ is simply the 
derivative of $F(x)$ with respect to its argument).
We define $R_\up{vir}$ such that
the mean density within $R_\up{vir}$ is $\Delta_\up{vir}\bar\rho_m$, and unless
otherwise stated we assume $\Delta_\up{vir}=200$.
We further split the one-halo term by discriminating
central and satellite galaxies (see, e.g., \citealt{andreas,yang,zheng2}),
\begin{equation}
\langle N\rangle_M\up{F}'(x)=\langle N_\up{cen}\rangle_M
\up{F}'_\up{cen}(x)+\langle N_\up{sat}\rangle_M
\up{F}'_\up{sat}(x).
\end{equation}
Pairs involving a central galaxy simply follow the radial mass profile
$\rho(r)$, so $\up{F}'_\up{cen}(x)\propto\rho(r)r^2$.
The distribution $\up{F}'_\up{sat}(x)$ of satellite galaxy-matter pairs is the
convolution of the galaxy and matter profiles. We assume a spherical NFW
profile \citep{nfw}, truncated at $R_\up{vir}$, for both dark matter and 
satellite galaxies. We compute the dark matter concentration parameter
$c_\up{dm}$ using the relation of \citet{james}. We allow the galaxy
concentration to be different, $c_\up{gal}=\alpha_c c_\up{dm}$, but adopt 
$\alpha_c=1$ as our standard assumption.

On scales much larger than the virial diameter of the largest halo, the
galaxy-matter correlation function is equal to $\xi_\up{mm}(r)$ multiplied by
a galaxy bias factor
\begin{equation}
b_\up{gal}={1\over {\bar n}_g}\int_0^\infty dM{dn\over dM}\langle N\rangle_M 
b_\up{h}(M),
\label{eq:largeR}
\end{equation}
where $b_\up{h}(M)$ is the bias factor of halos of mass $M$. However, an 
accurate calculation on intermediate scales must account for the finite extent
of halos, for the scale dependence of $b_\up{h}(M)$, and for halo exclusion ---
two spherical
halos cannot be separated by less than the sum of their virial radii.
It is convenient to do the calculation in Fourier space, where the convolutions
of halo profiles become multiplications of their Fourier transforms. Our 
complete series of expressions for the two-halo contribution to 
$\xi_\up{gm}(r)$ is 
\begin{equation}
1+\xi_\up{gm}^\up{2h}(r)=\left({\bar n_g'\over\bar n_g}\right)\left[1+
\xi_\up{gm}^\up{2h^\prime}(r)\right],
\label{eq:xi2h}
\end{equation}
where
\begin{equation}
\xi_\up{gm}^\up{2h^\prime}(r)={1\over2\pi^2}\int_0^\infty dk k^2 
P_\up{gm}^{2h^\prime}(k|r){\sin(kr)\over kr},
\end{equation}
is the Fourier transform of
\begin{eqnarray}
&&P_\up{gm}^\up{2h^\prime}(k|r)=P_m(k){1\over{\bar n}'_g}\int_0^\infty \!\!\!
dM_1{dn\over dM_1}\langle N\rangle_{M_1}b_\up{h}(M_1|r)~y_g(k,M_1) \nonumber\\ 
&&\times\int_0^\infty\!\!\!dM_2{dn\over dM_2}{M_2\over\bar\rho_m}b_\up{h}
(M_2|r)~y_m(k,M_2)p_\up{no}(x|M_1,M_2),
\label{eq:two}
\end{eqnarray}
where $y_\up{g}(k,M)$ and $y_\up{m}(k,M)$
are the normalized Fourier counterparts of the galaxy and the matter profiles,
and
\begin{equation}
\bar n_g^{\prime2}=\int_0^\infty\!\!\! dM_1 {dn\over dM_1} \langle N
\rangle_{M_1}\int_0^\infty\!\!\! dM_2 {dn\over dM_2} \langle 
N\rangle_{M_2}p_\up{no}(x|M_1,M_2).
\label{eq:ngave}
\end{equation}
In these expressions, $p_\up{no}(x|M_1,M_2)$ represents the probability that
two halos of mass $M_1$ and $M_2$ with scaled separation 
$x\equiv r/(R_\up{vir,1}+R_\up{vir,2})$ do not overlap. For spherical halos,
$p_\up{no}(x)$ would be a step function at $x=1$, but \citet{mtl} found that an
accurate separation of the 1-halo and 2-halo contributions to $\xi_\up{gg}(r)$
requires accounting for the non-spherical
shapes of halos identified by the FOF algorithm. We adopt their
expression, based on a fit to Monte Carlo realizations 
of ellipsoidal halo pairs
with a reasonable distribution of axis ratios:
$p_\up{no}(x)=0$ for $x<0.8$, $p_\up{no}(x)=1$ for 
$x>1.09$, and $p_\up{no}(x)=(3y^2-2y^3)$ with $y=(x-0.8)/0.29$ for 
$0.8\leq x\leq1.09$. The restricted number density $n_g^\prime$ is the mean
space density of galaxies residing in allowed (i.e., non-overlapping) halo
pairs at separation $r$. Since $p_\up{no}(x|M_1,M_2)$ and the halo bias factors
$b_\up{h}(M|r)$ depend on $r$, one must evaluate 
equations~(\ref{eq:xi2h})--(\ref{eq:ngave})
separately for each value of $r$ where one wants to know $\xi_\up{gm}(r)$.
The double integrals in equations~(\ref{eq:two}) and (\ref{eq:ngave}) are
non-separable because of the $M_1$-dependence of $p_\up{no}(x)$. 
In principle, one should separately compute equation~(\ref{eq:two}) for central
and satellite galaxies and sum the results, since $y_g(k,M)$ is different 
in the two cases, but we have tested and found that ignoring this subtlety has
negligible effect.

For scale-dependent halo bias factors, we adopt the expression
\begin{equation}
b_\up{h}^2(M|r)=b^2_\up{asym}(M)\times{\left[1+1.17\xi_\up{mm}(r)\right]^{1.49}
\over\left[1+0.69\xi_\up{mm}(r)\right]^{2.09}}~,
\label{eq:sbias}
\end{equation}
from \citet{mtl}. We also 
use Tinker~et~al.'s (2005a; Appendix A) expressions for
the asymptotic bias factors $b_\up{asym}(M)$. These follow the formulation of 
\citet{raby2}, but with different parameter values that yield a substantially
better fit to the simulations. For ``concordance'' cosmological parameters, 
these
bias factors are similar to those of \citet{uros2}, but they are more accurate
for models with different matter power spectra. We use Smith~et~al.'s (2003)
approximation for the non-linear power spectrum $P_\up{mm}(k)$ and correlation
function $\xi_\up{mm}(r)$ in equations~(\ref{eq:two}) and (\ref{eq:sbias}).
One could in principle use $P_\up{lin}(k)$ instead of $P_\up{mm}(k)$; this
would require a different (though still scale-dependent)
expression for $b_\up{h}(M|r)$ with a separate
$N$-body calibration.

The calculation as we have described it is a straightforward generalization
of the $\xi_\up{gg}(r)$ calculation presented by \citet{mtl}, 
who tested its accuracy over the range $\sigma_8=0.6-0.95$ for both the
$\Gamma=0.2$ simulations used here and for a similar set with $\Gamma=0.12$.
However, a
significant technical, and to some degree conceptual issue arises with the
evaluation of the second integral in equation~(\ref{eq:two}). Since we assign
all galaxy-matter pairs to either the 1-halo or 2-halo terms, we implicitly
assume that all dark matter is in halos of some mass, and thus
\begin{equation}
\int_0^\infty dM~{dn\over dM}~M =\bar\rho_m.
\label{eq:cons}
\end{equation}
More importantly for present purposes, the distribution of matter 
is by definition
unbiased with respect to itself, and therefore
\begin{equation}
\int_0^\infty dM~{dn\over dM}~{M\over\bar\rho_m}~b_\up{h}(M|r)= 1.
\label{eq:bi}
\end{equation}
The \citet{jenkins} mass function and \citet{mtl} halo bias factors used here
are fits to simulations over a finite range of halo masses, and they do not
satisfy either of these constraints. To impose the constraint 
(\ref{eq:bi}) explicitly, we break the second integral of 
equation~(\ref{eq:two}) at a halo mass $M_\up{brk}=10^8~\hmsun$ and evaluate it
as 
\begin{eqnarray}
&&\int_0^\infty\!\!\!dM_2{dn\over dM_2}{M_2\over\bar\rho_m}b_\up{h}(M_2|r)~
y_m(k,M_2)p_\up{no}(x|M_1,M_2)\nonumber \\
&&\approx
\int_{M_\up{brk}}^\infty\!\!\!dM_2{dn\over dM_2}{M_2\over\bar\rho_m}b_\up{h}
(M_2|r)~y_m(k,M_2) p_\up{no}(x|M_1,M_2) \nonumber \\ 
&&+ \left[1-\int_{M_\up{brk}}^\infty\!\!\! dM^\prime \frac{dn}{dM^\prime}
\frac{M^\prime}{\bar\rho_m}b_\up{h}(M'|r) \right].
\label{eq:new2h}
\end{eqnarray}
For the term in brackets on the right hand side, we make the (good) 
approximation that $p_\up{no}(x)=y(k,M)=1$ for halos with $M<M_\up{brk}$
at all radii of interest for our calculation,
then apply equation~(\ref{eq:bi}). We find that this procedure is necessary to
obtain accurate results. 
The same problem does not arise for integrals involving
$\langle N\rangle_M$ because the mean occupation itself goes to zero at low
halo masses.

We must make one further adjustment to the analytic model before testing it
against the populated $N$-body halos described in \S~\ref{sec:numerical}.
These halos are identified by the FOF algorithm, which, roughly speaking, 
selects particles within an isodensity surface. The {\it mean} overdensity
$\Delta_\up{vir}$ within this surface depends on the halo profiles. To 
compute the effective value of $\Delta_\up{vir}$ for our simulations, we 
calculate the mean density within spheres centered on the most-bound particles
of the FOF halos that enclose the halo's FOF mass. The canonical value of
$\Delta_\up{vir}=200$ is accurate for our central model with $\Omega_m=0.3$ and
$\sigma_8=0.8$.
However, the $\Delta_\up{vir}$
values for other models, listed in Table~\ref{tab:par}, deviate by up to 20\%.
The trend is as expected: halos in low $\Omega_m$ models are more concentrated
because they form earlier, and they have higher values of $\Delta_\up{vir}$.
However, the $\Omega_m$-dependence of FOF halos selected with constant linking
parameter is much weaker than the variation of
virial densities predicted by the spherical collapse model (e.g., 
\citealt{bryan}). When calculating virial radii as a function of halo mass,
we use the $\Delta_\up{vir}$ values in Table~\ref{tab:par}.

The attached bottom frame in Figure~\ref{fig:comp}a shows the fractional
difference between the analytic model and $N$-body results for the 
galaxy-galaxy correlation function, 
$(\xi_\up{analytic}-\xi_{N\mbox{-}\up{body}})/\xi_{N\mbox{-}\up{body}}$, 
for the five
cosmological/HOD models listed in Table~\ref{tab:par}. 
Error bars represent the statistical uncertainty in the mean value of 
$\xi_\up{N\mbox{-}\up{body}}$, computed from the dispersion among the five
independent simulations; for clarity, we show these only for the central
model with $\Omega_m=0.3$. The differences between the analytic and numerical
results are usually less than 5\% at all $r>0.1~\hmpc$. The one deviation
that is clearly statistically significant is the rapid turn-up
of the residuals at $r\simeq0.1~\hmpc$, which reflects the smoothing effect
of the simulation's gravitational force softening. The marginally significant,
$\sim5\%$ discrepancy at $r\simeq 0.8~\hmpc$ suggests that our ellipsoidal
exclusion correction still underestimates the number of close halo pairs, even
though it allows halos to be separated by less than the sum of their virial
radii. Without this correction, the deviation would fall off the
bottom of the plot (see Fig.~10 in \citealt{mtl}). The $\sim 8\%$
divergence of the models at $r\simeq 8~\hmpc$ results from the deviations of
the \citet{smith} non-linear matter power spectrum from our simulation results.
This difference could be partly an artifact of our finite box size, but the
systematic dependence on cosmological parameters and reconvergence of results
at $r=20~\hmpc$ suggest that it is mostly a result of slight non-universality
of the \citet{smith} formula, though it is hard to reach a definitive 
conclusion because of the substantial statistical uncertainties at these
scales.

The bottom frames in Figure~\ref{fig:comp}b and \ref{fig:comp}c show equivalent
fractional differences for $\xi_\up{gm}(r)$ and $\Delta\Sigma(r)$. The 
residuals for $\xi_\up{gm}(r)$ are similar to those for $\xi_\up{gg}(r)$, with
the rise at $r\simeq0.1~\hmpc$, again
reflecting the force softening in the numerical
simulation. Since $\Delta\Sigma(r)$ depends on $\xi_\up{gm}$ at all separations
less than $r$ (see eq.[\ref{eq:int}]), and the deviation between analytic and
$N$-body $\xi_\up{gm}$ grows at smaller separations, a comparison between the
pure analytic calculation and the measurement of $\Delta\Sigma(r)$ from the
simulation shows a substantial offset at $r\lesssim1~\hmpc$. However,
this offset reflects the limited resolution of the $N$-body simulation, not the
failure of the analytic model. (Similar deviations at small scale were found
by \citealt{mandel}.) 
To remove this numerical artifact from the comparison, we 
set $\Delta\Sigma$ for the analytic model equal to the $N$-body value at
$r=0.1~\hmpc$, then use the analytic calculation of $\xi_\up{gm}(r)$ to obtain
the surface density for $r>0.1~\hmpc$. With this correction, the analytic
model for $\Delta\Sigma(r)$ is accurate to 5\% or better for all five 
cosmological models at all radii $0.1~\hmpc\leq r\leq 20~\hmpc$.

\begin{figure*}[t]
\centerline{\epsfxsize=5.5truein\epsffile{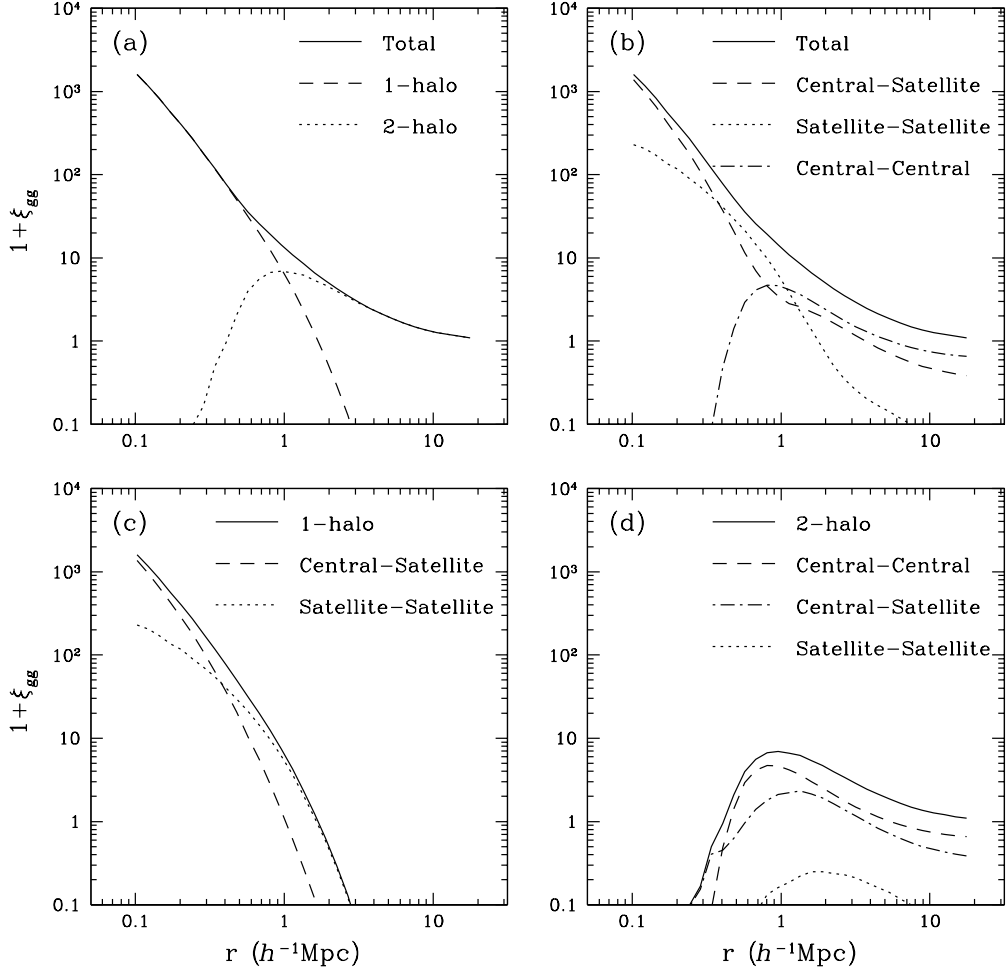}}
\caption{Dissection of the galaxy-galaxy correlations for the fiducial model. 
(a) Contributions of one-halo ($dashed$) and two-halo ($dotted$) galaxy pairs
to the full correlation function ($solid$). (b) Contributions of 
central-satellite ($dashed$), satellite-satellite ($dotted$), and 
central-central ($dot$-$dashed$) galaxy pairs. Panels (c) and (d) show the
central/satellite decompositions of the one- and two-halo terms individually.}
\label{fig:ggcs}
\end{figure*}

\begin{figure*}[t]
\centerline{\epsfxsize=5.5truein\epsffile{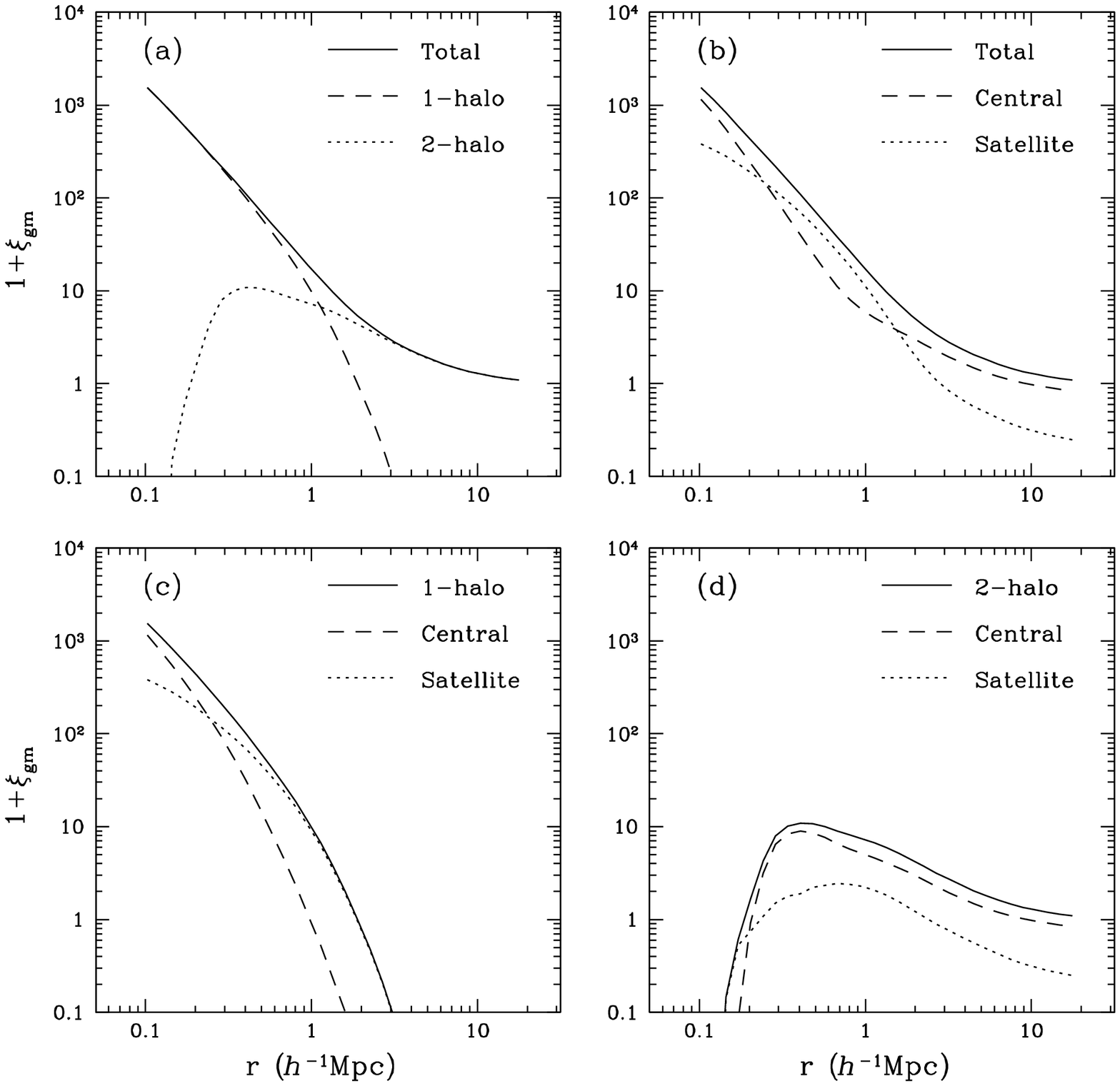}}
\caption{Dissection of the galaxy-matter correlations for the fiducial model,
in the same format as Fig.~\ref{fig:ggcs}. In panel (b)-(d), dashed and
dotted lines show galaxy-matter pairs involving a central galaxy or a
satellite galaxy, respectively.}
\label{fig:gmcs}
\end{figure*}

\subsection{Dissection of Correlation Functions}
\label{sec:dissection}
It is interesting to examine the separate contributions of central and 
satellite galaxies to the galaxy-galaxy and galaxy-matter correlation 
functions. Figure~\ref{fig:ggcs}a shows the familiar decomposition of
$\xi_\up{gg}(r)$ into 1-halo and 2-halo contributions (see \citealt{andreas}
for extensive discussion). We adopt the fiducial model with $\Omega_m=0.3$,
$\sigma_8=0.8$, and the corresponding HOD parameters listed in 
Table~\ref{tab:par}, and we use the $N$-body simulation measurements. 
The 1-halo 
term dominates at small scales, but it drops rapidly towards larger $r$ as 
halos with large virial diameters become increasingly rare. At large scales,
the 2-halo term is a linearly biased version of the matter correlation 
function, but at small scales it turns over and drops because of halo 
exclusion. We plot $1+\xi(r)$ rather than $\xi(r)$ 
itself because pair counts are
additive, so individual contributions sum to give the total $1+\xi(r)$; this
consideration becomes especially important for the central-satellite 
decompositions shown in subsequent panels. The transition between 1-halo and
2-halo dominance occurs at roughly the virial diameter of $M_\star$ halos,
where $M_\star$ is the characteristic scale of the halo mass function.

Figure~\ref{fig:ggcs}b separates $1+\xi_\up{gg}(r)$ into central and satellite
galaxy contributions. Central-satellite pairs dominate
$\xi_\up{gg}(r)$ at $r\lesssim0.4~\hmpc$, and central-central pairs at
$r\gtrsim1~\hmpc$, while satellite-satellite
pairs dominate by a small factor in the intermediate regime.
Figure~\ref{fig:ggcs}c and \ref{fig:ggcs}d show the separate central and
satellite contributions to the 1-halo and 2-halo terms, revealing the origin
of the behavior in Figure~\ref{fig:ggcs}b. Central-satellite
pairs dominate $\xi^\up{1h}(r)$ 
at small scales because in this regime most pairs come from the
most common halos that are large enough to host a galaxy pair. These halos
have $\langle N\rangle_M<3$, and they therefore have more central-satellite 
galaxy pairs
than pairs that involve only satellites. Conversely, satellite-satellite
pairs dominate $\xi^\up{1h}(r)$ at large scales because the halos with virial 
diameters large
enough to host pairs at these separations have $\langle N\rangle_M>3$.
Finally, pairs involving at least one central galaxy
dominate the 2-halo term by a large factor at all
separations, because only halo pairs in which {\it both} halos have 
$\langle N\rangle_M>2$ contribute, on average, more satellite-satellite
pairs than central-central or central-satellite pairs.
Such halos are much less common than those with
$1\leq\langle N\rangle_M\leq2$. Thus, satellite-satellite
pairs make a major contribution
to the total $\xi_\up{gg}(r)$ only over the radial range in which 1-halo 
contributions from high mass halos are dominant.

Figure~\ref{fig:gmcs} shows the equivalent dissection of $\xi_\up{gm}(r)$. The 
overall behavior is very similar to that seen in Figure~\ref{fig:ggcs}.
The 2-halo term extends to somewhat smaller scales because halos with mass
near $M_\up{min}$ can still
form galaxy-matter pairs with lower mass halos that have
smaller virial radii. The satellite contribution to the 2-halo term 
is analogous to the sum of 
central-satellite and satellite-satellite pairs for $\xi_\up{gg}(r)$
because there are no ``central'' dark matter particles.
(The normalization is lower than in Figure~\ref{fig:sub} because we now 
calculate the expected number of galaxy-matter pairs using all galaxies instead
of satellites alone.) However, it is still a factor of 
3$-$5 below the central 2-halo term at nearly all separations. From 
Figure~\ref{fig:gmcs}, we can understand why the individual halos of satellite
galaxies appear to have so little impact on the SPH results discussed in
\S~\ref{sec:comp}. Satellite galaxies dominate the 1-halo term of 
$\xi_\up{gm}(r)$ only beyond $r\simeq0.25~\hmpc$. However, the individual halos
of satellites orbiting in larger groups are usually tidally truncated well
inside this radius, on scales where the signal is swamped by the contribution
from central galaxies. 
Note that the satellite fraction of galaxy samples is 
typically less than 30\%, and hence the lensing signal of satellite galaxies is
smaller by an order of magnitude than that of central galaxies.
The dark halos of satellites in groups and clusters
can be measured by galaxy-galaxy lensing \citep{priya},
but only by first
identifying satellites and measuring $\Delta\Sigma(r)$ for them specifically.

\section{From Galaxy-Galaxy Lensing to Cosmological Parameters}
\label{sec:dep}
Having established the accuracy of the analytic model, we can now use it to
investigate the dependence of $\Delta\Sigma(r)$ on $\Omega_m$ and $\sigma_8$.
We consider a well-defined sample of galaxies, choose HOD parameters for
each ($\Omega_m$, $\sigma_8$) combination by fitting the mean space density and
projected correlation function of this sample, then predict $\Delta\Sigma(r)$.
In this section, we focus on the sample of SDSS galaxies with $M_r\leq-21$,
with the projected correlation function, error covariance matrix, and mean
space density $\bar{n}_g=1.17\times10^{-3}~h^3\mpc^{-3}$ taken 
from \citet{idit3}.
We also present predictions for a fainter luminosity threshold, $M_r\leq-20$,
again using \citet{idit3}'s observational constraints.\footnote{Specifically,
we use the \citet{idit3} measurements for the $M_r\leq-20$ sample with limiting
redshift $z=0.06$.} At large scales, we
expect to recover the linear theory, linear bias result, 
$\Delta\Sigma\propto\Omega_m/b\propto\sigma_8\Omega_m$, but we can extend
the predictions to intermediate and small scales using the full analytic
model.

We make two significant changes in our application of the analytic model.
First, we use a {\scriptsize CMBFAST}
transfer function \citep{cmbfast}, computed for
$\Omega_m=0.3$, $h=0.7$, $\Omega_b=0.04$, in place of the \citet{george}
parametrization adopted in our $N$-body simulations. This change to the
transfer function has little effect on the HOD parameters inferred by fitting
$w_p(r_p)$, but it has a noticeable effect on the $\chi^2$ values of these 
fits, and it affects the $\Delta\Sigma(r)$ predictions themselves. Note that
we do $not$ change the transfer function when changing $\Omega_m$ from our 
fiducial value of 0.3; because the power spectrum shape is empirically well
constrained, we assume that any effect of changing $\Omega_m$ will be 
compensated by adjusting $h$, $\Omega_b$, or the inflationary index $n$
(which we set to one).
Second, we define halo virial radii assuming
$\Delta_\up{vir}=200$ for all $\Omega_m$, instead of the varying 
$\Delta_\up{vir}$ values listed in Table~\ref{tab:par} and used in 
\S~\ref{sec:analytic}. This change simply
amounts to a slight change in the halo definition; to identify these halos
in $N$-body simulations, one would need to adjust the FOF linking length 
slightly with $\Omega_m$. The \citet{james} concentration parameters are 
defined for different ($\Omega_m$-dependent) $\Delta_\up{vir}$ values, but
we rescale them to our $\Delta_\up{vir}$ definition. We still adopt the
\citet{jenkins} halo mass function for all cosmological models, with no 
rescaling.

\begin{figure*}[t]
\centerline{\epsfxsize=6.8truein\epsffile{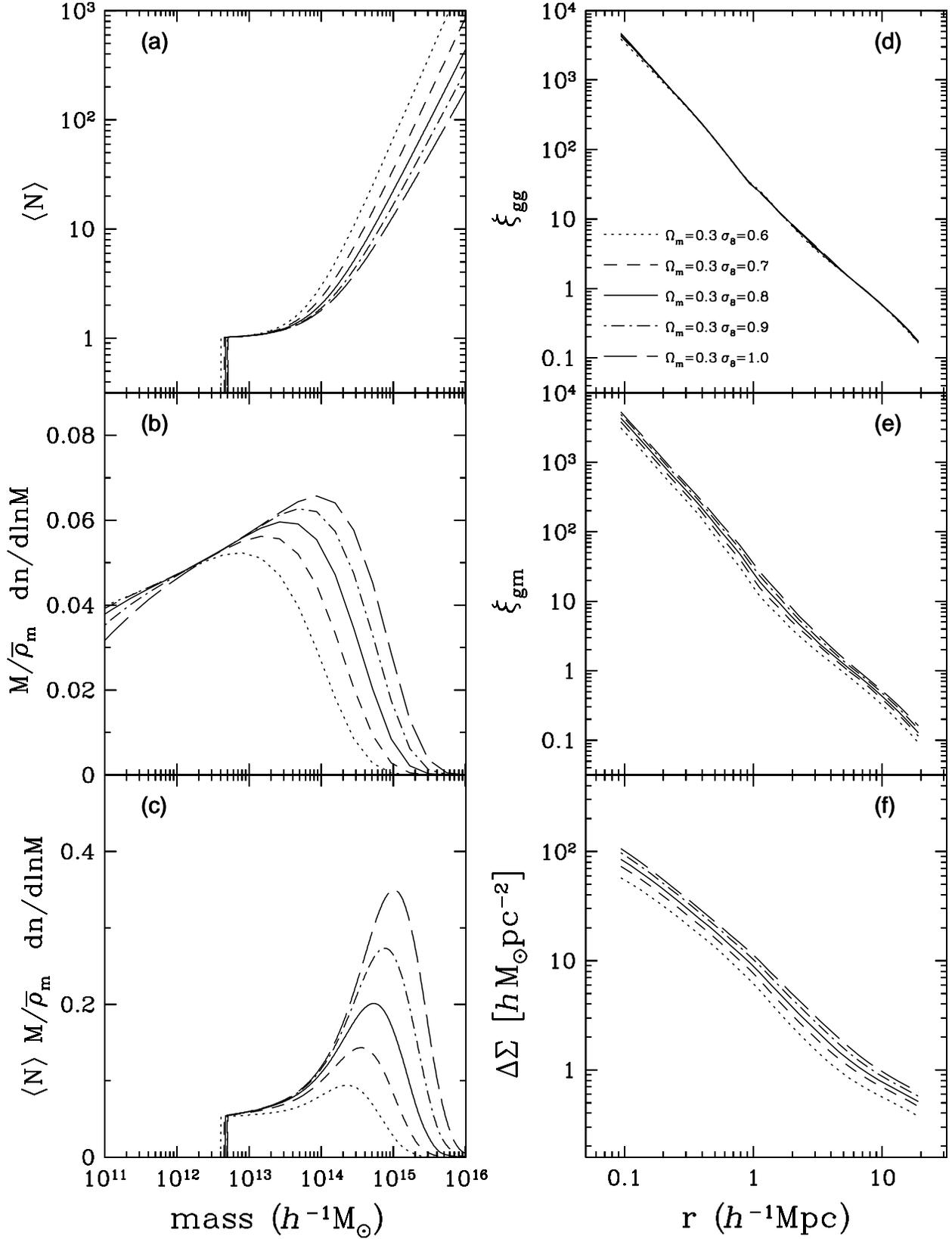}}
\caption{Clustering contributions and clustering signals for the model 
sequence with fixed $\Omega_m$ and varying $\sigma_8$. (a) Mean halo occupation
functions, determined by fitting the $w_p(r_p)$ data, with $\sigma_8$ 
increasing from top to bottom (see panel~d legend). 
(b) Fraction of matter per
logarithmic bin of halo mass. (c) Fraction of galaxy-matter pairs per 
logarithmic bin of halo mass. Note that pairs in higher mass halos are spread
over a wider range of separations, diluting the contribution to any given 
$r\rightarrow r+dr$ bin. Panel~(d), (e), and (f) show the galaxy-galaxy
correlation function, galaxy-matter correlation function, and excess surface
density, respectively.}
\label{fig:sig8}
\end{figure*}

Figure~\ref{fig:sig8} illustrates the results for a sequence of models with
$\sigma_8$ ranging from 0.6 to 1.0, all for $\Omega_m=0.3$. The HOD parameter
values required to match the \citet{idit3} $w_p(r_p)$ 
measurements are 
listed in Table~\ref{tab:par2}, and the mean occupation functions
$\langle N\rangle_M$ are shown in panel~\ref{fig:sig8}a. For lower $\sigma_8$,
matching the observed clustering requires a larger fraction of galaxies in
more massive, more biased halos, hence higher values of the
$\langle N_\up{sat}\rangle_M$ slope $\alpha_\up{sat}$. The resulting
galaxy correlation functions are very similar for all five values of 
$\sigma_8$, as shown in Figure~\ref{fig:sig8}d. Figure~\ref{fig:sig8}b plots
$(M/\bar\rho_m)~dn/d\ln M$, the fraction of mass contained in a (natural) 
logarithmic bin centered at mass $M$. For $\sigma_8=0.6$, this function peaks
near $M\sim10^{13}~\hmsun$, while for $\sigma_8=1.0$ it peaks near 
$M\sim10^{14}~\hmsun$. Figure~\ref{fig:sig8}c plots the same function 
multiplied by $\langle N\rangle_M$, a product that is 
proportional to the number of
1-halo galaxy-matter $pairs$ that arise in halos of mass $M$. This function
peaks at $M\sim3\times10^{14}~\hmsun$ for $\sigma_8=0.6$ and 
$M\sim10^{15}~\hmsun$ for $\sigma_8=1.0$. The trend of $\alpha_\up{sat}$ with
$\sigma_8$ partly compensates the trend of the
mass distribution in Figure~\ref{fig:sig8}b, reducing the order-of-magnitude
shift in the peak location to a factor of three. 
While high mass halos near the peak contribute a substantial fraction of
all galaxy-matter pairs, these pairs are spread over a larger projected area,
so the contribution in a given $r\sim r+dr$ bin is multiplied by an additional
factor that scales roughly as $R_\up{vir}^{-2}\sim M^{-2/3}$.
For all values of $\sigma_8$,
the fraction of 1-halo galaxy-matter pairs is tiny for 
$M\geq5\times10^{15}~\hmsun$.

\begin{deluxetable}{cccccc}
\tablewidth{0pt}
\tablecaption{HOD Parameters for Different $\sigma_8$ Models}
\tablehead{\colhead{Model} & \colhead{$\Omega_m$} & \colhead{$\sigma_8$}
& \colhead{$M_\up{min}(\hmsun)$} & \colhead{$M_1(\hmsun)$}
& \colhead{$\alpha_\up{sat}$}}
\startdata
1 & 0.3 & 0.6 & $4.04\times10^{12}$ & $6.28\times10^{13}$ & 1.52 \\
2 & 0.3 & 0.7 & $4.46\times10^{12}$ & $7.98\times10^{13}$ & 1.40 \\
3 & 0.3 & 0.8 & $4.71\times10^{12}$ & $9.58\times10^{13}$ & 1.31 \\
4 & 0.3 & 0.9 & $4.85\times10^{12}$ & $1.11\times10^{14}$ & 1.25 \\
5 & 0.3 & 1.0 & $4.95\times10^{12}$ & $1.23\times10^{14}$ & 1.19 \\
\enddata
\tablecomments{The HOD parameters are chosen to reproduce the
same clustering of the SDSS galaxy sample of $M_r\leq-21$ and to match the 
number density of galaxies $\bar n_g=1.17\times10^{-3}~(\hmpc)^{-3}$.}
\label{tab:par2}
\end{deluxetable}

Figures~\ref{fig:sig8}e and \ref{fig:sig8}f show the galaxy-matter correlation
functions and excess surface density profiles, respectively, 
for this model sequence.
At large scales, $\xi_\up{gm}(r)$ and $\Delta\Sigma(r)$ increase with 
$\sigma_8\propto1/b$ as expected from linear theory. A similar increase appears
on small scales because of the larger fraction of galaxy-matter pairs in more
massive halos. In fact, the shapes of $\xi_\up{gm}(r)$ and $\Delta\Sigma(r)$
appear remarkably constant over the full range $0.1~\hmpc\leq r\leq20~\hmpc$,
a point we quantify below.

\begin{figure*}[t]
\centerline{\epsfxsize=7truein\epsffile{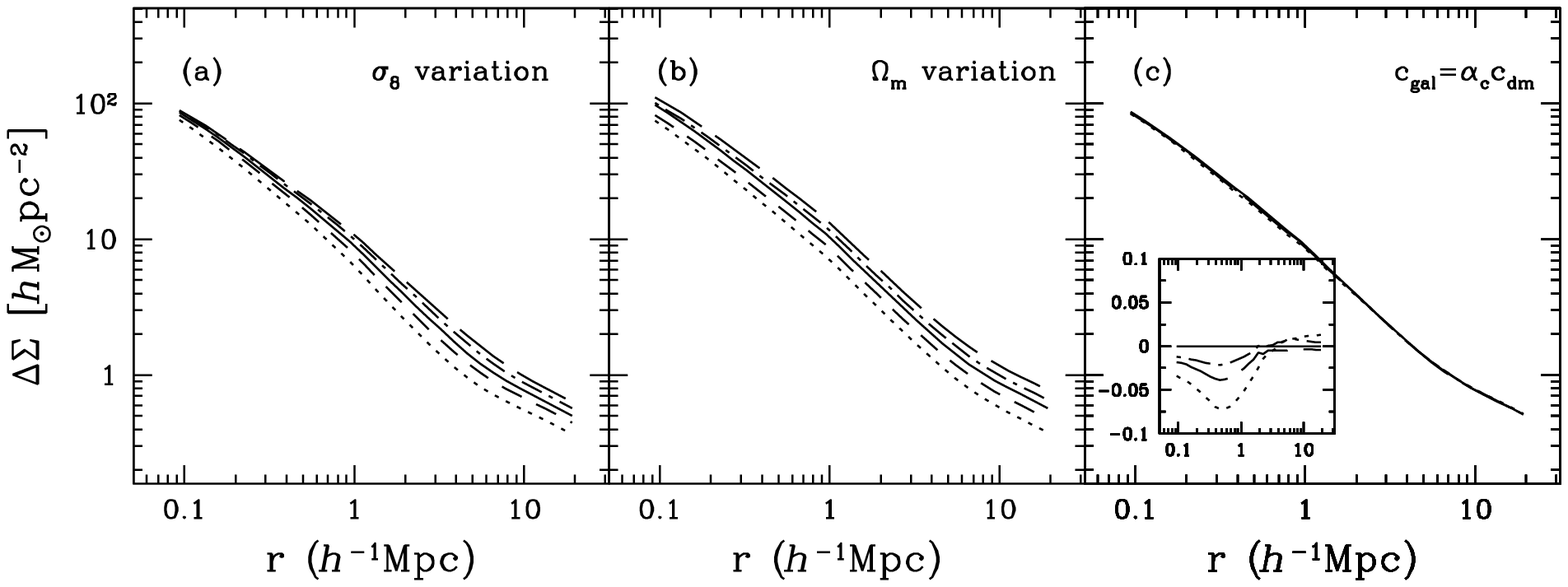}}
\caption{Excess surface density profiles for other model sequence.
We consider (a) the same model sequence as in Fig.~\ref{fig:sig8}, but with
halo concentrations held fixed at the values for $\Omega_m=0.3$ and 
$\sigma_8=0.8$, (b) a sequence of models with fixed $\sigma_8=0.8$ and 
$\Omega_m$ ranging from 0.2 ({\it dotted}) to 0.4 ({\it long dashed}) in steps
of 0.05, and (c) models with $\Omega_m=0.3$ and $\sigma_8=0.8$ in which
galaxy profile concentrations satisfy $c_\up{gal}=\alpha_c c_\up{dm}$, with 
$\alpha_c=0.3~(dotted)$, 0.5 ($long$-$dashed$), 0.7 ($dot$-$dashed$), and
1.0 $(solid)$. The inset panel shows fractional deviations from the 
$\alpha_c=1.0$ model.}
\label{fig:ana}
\end{figure*}

Figure~\ref{fig:ana} shows $\Delta\Sigma(r)$ for three model sequences with
different variations. In Figure~\ref{fig:ana}a, we consider the same sequence
of increasing $\sigma_8$, fixed $\Omega_m$ shown in
Figure~\ref{fig:sig8}, but we
always keep halo concentrations fixed at the values predicted for 
$\sigma_8=0.8$. We adjust HOD parameters slightly from the values listed
in Table~\ref{tab:par2} to obtain the minimum-$\chi^2$ fit to the projected 
correlation function with the new halo concentrations.
In the large $r$, 2-halo regime, $\Delta\Sigma(r)$ is nearly
identical to that shown in Figure~\ref{fig:sig8}f.
However, fixing the concentrations to those of the central model has an
important effect at small scales, causing the $\Delta\Sigma(r)$ curves to
converge. The constancy of shape in Figure~\ref{fig:sig8}f is thus partly
a consequence of the changes in halo concentrations in different 
$\sigma_8$ models; higher $\sigma_8$ leads to earlier halo collapse and higher
concentration, boosting $\Delta\Sigma(r)$.

Figure~\ref{fig:ana}b shows a sequence with fixed $\sigma_8=0.8$ and $\Omega_m$
varying from 0.2 to 0.4 in steps of 0.05. For this sequence, the HOD 
parameters $M_\up{min}$ and $M_1$ scale in proportion to $\Omega_m$,
though we again make slight adjustments to fit $w_p(r_p)$.
If halo concentrations were independent of $\Omega_m$, and we did not make
those small adjustments, then $\xi_\up{gg}(r)$ and
$\xi_\up{gm}(r)$ would be identical for all five models after this 
mass rescaling,
since the halo mass function and halo bias factors are functions of $M/M_\star$
and $M_\star\propto\Omega_m$ (see \citealt{zheng1} for further discussion).
In this case, $\Delta\Sigma(r)$ would have a constant shape and an amplitude
proportional to $\Omega_m$. Figure~\ref{fig:ana}b shows roughly this behavior,
but the trend of higher concentration for lower $\Omega_m$ produces a weak
convergence of models at small $r$.

We have so far assumed that satellite galaxies trace the dark matter in halos,
with the same NFW radial profile. We now relax this assumption and allow the
satellite profiles to have a lower concentration parameter, as suggested by
some numerical studies (see \S~\ref{sec:comp} and \citealt{nagai}). 
Figure~\ref{fig:ana}c compares models with $\Omega_m=0.3$, $\sigma_8=0.8$,
and satellite concentration parameters $c_\up{gal}=\alpha_c c_\up{dm}$ with
$\alpha_c=0.3$, 0.5, 0.7, and 1.0. We again adjust HOD parameters to
fit $w_p(r_p)$ after changing galaxy concentrations.
These adjustments partly compensate for the changes in galaxy concentration,
so the effect of a radial profile change is somewhat smaller here than
in \S~\ref{sec:comp} (Fig.~\ref{fig:sph}), where we kept other HOD parameters
fixed. For $\alpha_c\geq0.7$, the 
impact on $\Delta\Sigma(r)$ is under 3\% at all $r$. For $\alpha_c=0.3$, the
effect rises to 7\% at the scale $r\sim0.5-1~\hmpc$ where satellite galaxies
make a dominant contribution to $\xi_\up{gm}(r)$ (see Fig.~\ref{fig:gmcs}b).
Small differences between galaxy and dark matter concentrations can thus be
safely neglected, but large differences can have a small but measurable impact
at $r<2~\hmpc$.

\begin{deluxetable}{rcrrrcrr}
\tabletypesize{\scriptsize}
\tablewidth{0pt}
\tablecaption{The Weak Lensing Signal}
\tablehead{ \colhead{} & \multicolumn{3}{c}{$M_r\leq-21$} & \colhead{} &
\multicolumn{3}{c}{$M_r\leq-20$} \\ \cline{2-4} \cline{6-8}
\colhead{$\log r$} & \colhead{$\log\Delta\Sigma(r)$} & 
\colhead{$\alpha$} & \colhead{$\beta$} & \colhead{} &
\colhead{$\log\Delta\Sigma(r)$} & \colhead{$\alpha$} & \colhead{$\beta$} }
\startdata
$-$1.029 &    1.929 & 0.65 & 1.13 & &   1.582 & 0.71 & 0.84 \\
$-$0.952 &    1.870 & 0.68 & 1.10 & &   1.513 & 0.73 & 0.83 \\
$-$0.875 &    1.808 & 0.71 & 1.08 & &   1.446 & 0.75 & 0.83 \\
$-$0.798 &    1.743 & 0.73 & 1.06 & &   1.378 & 0.77 & 0.83 \\
$-$0.721 &    1.674 & 0.76 & 1.04 & &   1.310 & 0.79 & 0.85 \\
$-$0.643 &    1.602 & 0.79 & 1.03 & &   1.244 & 0.81 & 0.85 \\
$-$0.567 &    1.529 & 0.80 & 1.03 & &   1.181 & 0.83 & 0.87 \\
$-$0.489 &    1.453 & 0.83 & 1.03 & &   1.116 & 0.84 & 0.91 \\
$-$0.412 &    1.379 & 0.85 & 1.02 & &   1.050 & 0.86 & 0.96 \\
$-$0.335 &    1.306 & 0.87 & 1.03 & &   0.982 & 0.88 & 1.00 \\
$-$0.258 &    1.227 & 0.89 & 1.06 & &   0.922 & 0.89 & 1.08 \\
$-$0.181 &    1.145 & 0.91 & 1.10 & &   0.860 & 0.91 & 1.13 \\
$-$0.104 &    1.064 & 0.92 & 1.15 & &   0.806 & 0.92 & 1.16 \\
$-$0.027 &    0.983 & 0.93 & 1.19 & &   0.753 & 0.95 & 1.19 \\
0.050    &    0.891 & 0.94 & 1.24 & &   0.686 & 0.96 & 1.24 \\
0.127    &    0.793 & 0.95 & 1.31 & &   0.610 & 0.98 & 1.31 \\
0.204    &    0.697 & 0.96 & 1.37 & &   0.535 & 0.99 & 1.38 \\
0.281    &    0.604 & 0.98 & 1.41 & &   0.456 & 1.00 & 1.44 \\
0.358    &    0.511 & 0.99 & 1.45 & &   0.378 & 1.01 & 1.49 \\
0.435    &    0.419 & 1.00 & 1.48 & &   0.297 & 1.02 & 1.53 \\
0.512    &    0.330 & 1.00 & 1.49 & &   0.218 & 1.03 & 1.55 \\
0.589    &    0.244 & 1.01 & 1.47 & &   0.137 & 1.04 & 1.55 \\
0.666    &    0.162 & 1.02 & 1.43 & &   0.059 & 1.04 & 1.51 \\
0.743    &    0.089 & 1.03 & 1.37 & &$-$0.013 & 1.05 & 1.43 \\
0.820    &    0.021 & 1.04 & 1.29 & &$-$0.076 & 1.05 & 1.35 \\
0.897    & $-$0.039 & 1.03 & 1.21 & &$-$0.131 & 1.06 & 1.27 \\ 
0.974    & $-$0.094 & 1.05 & 1.14 & &$-$0.196 & 1.05 & 1.20 \\
1.051    & $-$0.145 & 1.05 & 1.09 & &$-$0.242 & 1.05 & 1.15 \\
1.128    & $-$0.193 & 1.05 & 1.06 & &$-$0.288 & 1.05 & 1.12 \\
1.205    & $-$0.241 & 1.04 & 1.05 & &$-$0.333 & 1.04 & 1.12 \\
1.282    & $-$0.292 & 1.04 & 1.05 & &$-$0.382 & 1.04 & 1.12 \\
\enddata
\tablecomments{The excess surface densities $\Delta\Sigma_\up{FID}$ 
of the fiducial model ($\Omega_m=0.3$, $\sigma_8=0.8$) for SDSS galaxy
samples of $M_r\leq-21$ and $M_r\leq-20$.
The projected radius $r$ is in $\hmpc$, and $\Delta\Sigma$ is in 
$h\msun\up{pc}^{-2}$.}
\label{tab:delta}
\end{deluxetable}

\begin{figure*}[t]
\centerline{\epsfxsize=5.5truein\epsffile{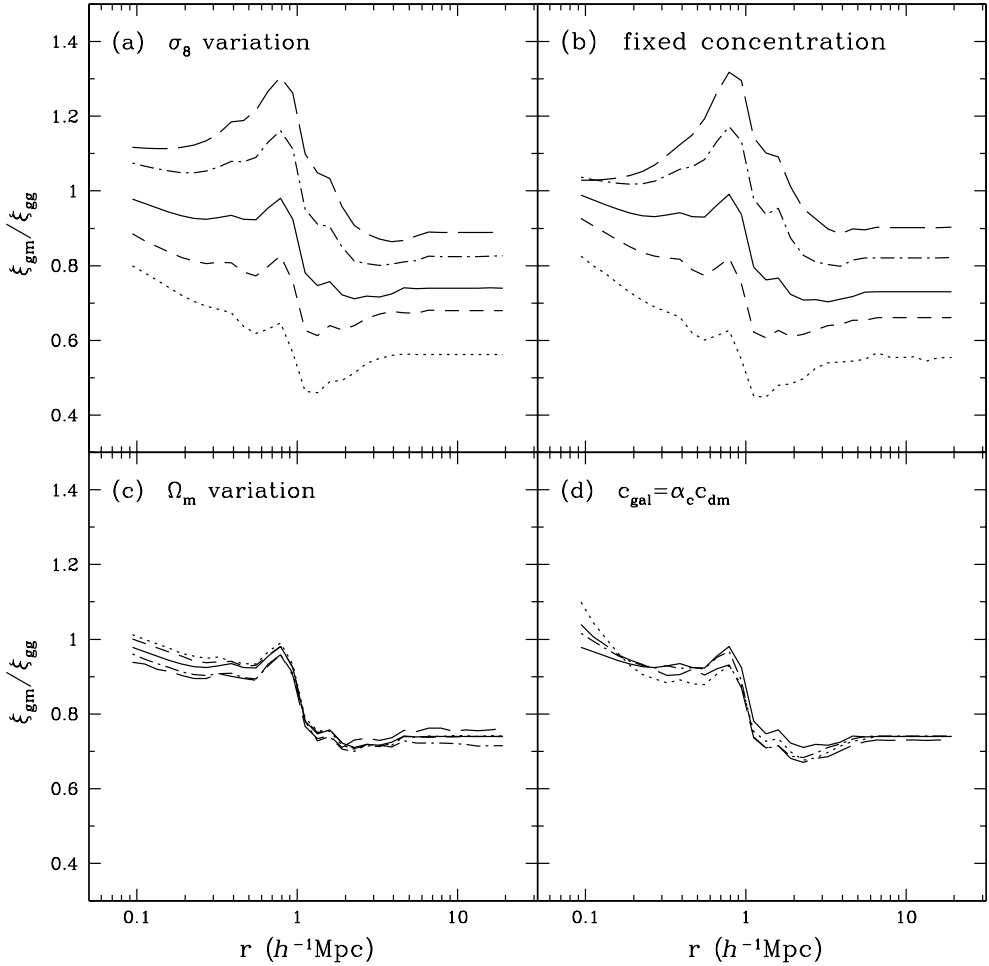}}
\caption{The ratio $\xi_\up{gm}(r)/\xi_\up{gg}(r)$, which is equal to 
$r_\up{gm}/b$ in the linear bias model. Panels~(a)-(d) show the four model
sequences illustrated in Figs.~\ref{fig:sig8} and \ref{fig:ana}. Line types
follow the same sequence as in those Figures, with $\sigma_8$ increasing from
0.6~($dotted$) to 1.0~($long$-$dashed$) in panels (a) and (b), $\Omega_m$
increasing from 0.2~($dotted$) to 0.4~($long$-$dashed$) in (c), and $\alpha_c$
increasing from 0.3~($dotted$) to 1.0~($solid$) in (d).}
\label{fig:rob}
\end{figure*}

Observational studies of galaxy-galaxy lensing often examine the ratio
$\xi_\up{gm}(r)/\xi_\up{gg}(r)$, which is equal to $r_\up{gm}/b$ in the linear
bias model (e.g., \citealt{henk3,henk2,erin}).
Figure~\ref{fig:rob} plots this ratio for the four model sequences shown
in Figures~\ref{fig:sig8} and \ref{fig:ana}. Figure~\ref{fig:rob}a shows the
fixed-$\Omega_m$, increasing $\sigma_8$ sequence of Figure~\ref{fig:sig8}.
Since HOD parameters in each model are adjusted to match the projected 
correlation, model differences are driven almost entirely by changes in 
$\xi_\up{gm}(r)$. At large scales, the ratio $\xi_\up{gm}(r)/\xi_\up{gg}(r)$ is
constant as predicted for linear bias, and $\xi_\up{gm}(r)$ increases in
proportion to $\sigma_8\propto1/b$. Comparison to the bias factor defined by
$b^2=\xi_\up{gg}(r)/\xi_\up{mm}(r)$ implies a cross-correlation coefficient
$r_\up{gm}\simeq0.9$ for all five models. However, in every case 
$\xi_\up{gm}(r)/\xi_\up{gg}(r)$ rises sharply at a scale $r\sim1~\hmpc$ near
the 1-halo to 2-halo transition. For $\xi_\up{gg}(r)$, this transition is
fairly sharp, producing measurable deviations from a power-law
\citep{andreas,idit2}. These deviations are smoothed out in $\xi_\up{gm}(r)$
because the contribution of halos below $M_\up{min}$ allows the 2-halo term
to overlap more with the 1-halo term (compare Figs.~\ref{fig:ggcs}a and
\ref{fig:gmcs}a), and the ratio $\xi_\up{gm}(r)/\xi_\up{gg}(r)$ therefore
shows a sharp feature reflecting the break in $\xi_\up{gg}(r)$. For higher
$\sigma_8$, the 1-halo term extends to larger $r$, and the jump in
$\xi_\up{gm}(r)/\xi_\up{gg}(r)$ is larger in amplitude but spread over a 
larger range of $r$. The break in $\xi_\up{gg}(r)$ is generally stronger
for more strongly clustered galaxy samples \citep{andreas,idit3}, and we
expect similar sample dependence for the $\xi_\up{gm}(r)/\xi_\up{gg}(r)$
jump.

\begin{figure*}[t]
\centerline{\epsfxsize=5.5truein\epsffile{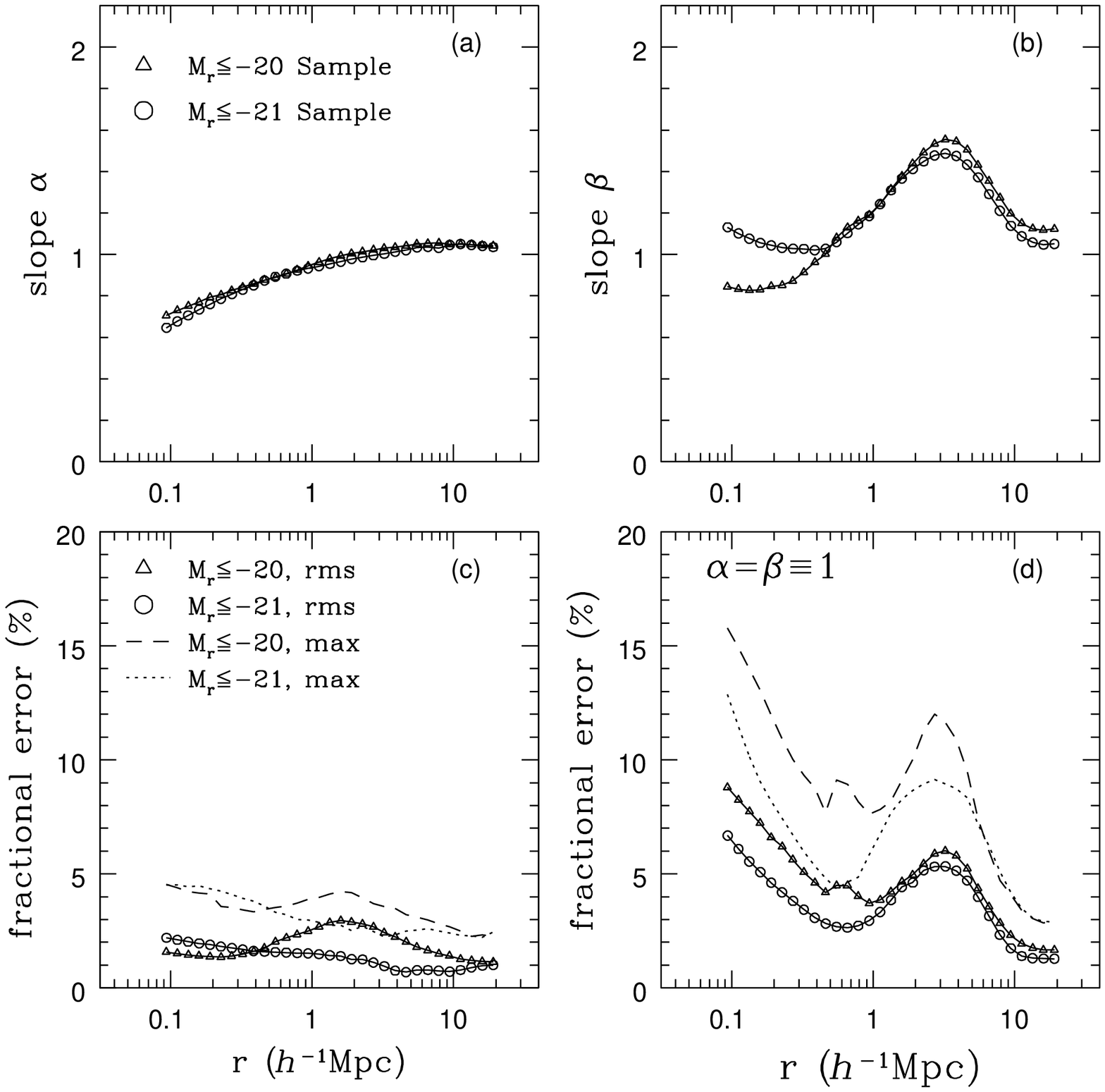}}
\caption{Parameters $\alpha(r)$ and $\beta(r)$ of the bias scaling relation
$\Delta\Sigma(r)\propto\Omega_m^\alpha\sigma_8^\beta$ (eq.[\ref{eq:rel}]) for 
samples matched to the SDSS $w_p(r_p)$ measurements of $M_r\leq-20$ 
($triangles$) and $M_r\leq-21$ ($circles$) galaxies. Panel~(c) shows the rms
and maximum fractional errors of this scaling relation, relative to the full
analytic calculation, over a model grid with $\Omega_m$ varying from 0.15 to
0.45 and $\sigma_8$ varying from $\sigma_8=0.6$ to 1.0. Panel~(d) shows the
same errors for linear bias scaling $\alpha=\beta=1$. In all cases the maximum
error arises for $\Omega_m=0.45$, $\sigma_8=1.0$.}
\label{fig:slope}
\end{figure*}

At small scales, $\xi_\up{gm}(r)/\xi_\up{gg}(r)$ is again roughly flat, but
at a level higher than the large scale ratio. As noted earlier, fixing halo
concentrations causes the galaxy-matter correlation function of different
$\sigma_8$ models to converge (Fig.~\ref{fig:ana}a), so the 
$\xi_\up{gm}(r)/\xi_\up{gg}(r)$ ratios also converge in this case
(Fig.~\ref{fig:rob}b). Figures~\ref{fig:rob}c and \ref{fig:rob}d show that
the effects of $\Omega_m$ or $c_\up{gal}$ variations are much smaller than 
those of $\sigma_8$ variations. The $\sim5\%$ model-to-model differences at 
small scales arise from their different halo
concentrations, while the smaller differences
at large scales reflect the slight changes in HOD parameters required to match
$w_p(r_p)$. Present observations \citep{henk2,erin} are consistent with
$\xi_\up{gm}(r)/\xi_\up{gg}(r)$ that is approximately scale-independent, but
the uncertainties are still fairly large, and testing for the feature predicted
in Figure~\ref{fig:rob} will require more careful replication of observational
procedures.

Figure~\ref{fig:rob} shows that the linear bias expectation of constant
$\xi_\up{gm}(r)/\xi_\up{gg}(r)$  
holds accurately for $r\geq4~\hmpc$
but fails at the 20$-$50\% level in the non-linear regime.
We can also ask how well the linear bias prediction 
$\Delta\Sigma\propto\Omega_m\sigma_8$ describes the {$scaling$} of 
$\Delta\Sigma(r)$ with cosmological parameters. To answer this question, and
to allow
easy scaling of our predictions with cosmological parameters, we adopt the
more general formula
\begin{equation}
{\Delta\Sigma(r)\over\Delta\Sigma_\up{FID}(r)}=\left({\Omega_m\over0.3}
\right)^\alpha\left({\sigma_8\over0.8}\right)^\beta
\label{eq:rel}
\end{equation}
and determine best-fit values of $\alpha$ and $\beta$ at each separation $r$.
Here $\Delta\Sigma_\up{FID}(r)$ is the excess surface density prediction
of the fiducial model with $\Omega_m=0.3$ and $\sigma_8=0.8$, and we fit 
$\alpha$
and $\beta$ using a full grid of models with $\sigma_8=0.6$, 0.7, 0.8, 0.9, 1.0
and $\Omega_m=0.15$, 0.2, 0.25, 0.3, 0.35, 0.4, 0.45. We assume 
$c_\up{gal}=c_\up{dm}$ in all cases.

Figures~\ref{fig:slope}a and \ref{fig:slope}b
plot the fitted values of $\alpha$ and $\beta$, respectively, as a function
of $r$. Results for $M_r\leq-20$ galaxies and $M_r\leq-21$ galaxies are 
similar, though the underlying $\Delta\Sigma_\up{FID}(r)$ is different in
the two cases. There are two notable departures from the linear bias values
$\alpha=\beta=1$. At small scales, $\alpha$ falls below one, reflecting the
weak convergence of $\Delta\Sigma(r)$ curves seen in Figure~\ref{fig:ana}b.
This convergence in turn reflects the $\Omega_m$-dependence of halo 
concentrations. At scales $r\sim2-5~\hmpc$, $\beta$ rises above unity,
corresponding to the slight divergence of $\Delta\Sigma(r)$ curves at these
scales in Figure~\ref{fig:sig8}f. Figure~\ref{fig:slope}c
shows the rms and maximum 
fractional errors between the $\Omega_m$ or $\sigma_8$
dependences predicted by the full analytic model and the scaling relation
(\ref{eq:rel}), calculated over our full model grid. The rms errors range
from $\sim1\%$ at large $r$ to $\sim3\%$ at intermediate $r$. 
The largest errors arise
for the $\Omega_m=0.45$, $\sigma_8=1.0$ model, and they are roughly twice
the rms errors. Figure~\ref{fig:slope}d shows
the result of adopting the linear bias scaling $\alpha=\beta=1$ in 
equation~(\ref{eq:rel}). The linear bias predictions are accurate at the
$\sim1\%$ (rms) to $\sim3\%$ (maximum error) level for $r\geq10~\hmpc$,
but the deviations become substantial at smaller $r$, with errors of 
$\sim5-16\%$ at $r=3~\hmpc$ and $r=0.1~\hmpc$ for the $\Omega_m=0.45$, 
$\sigma_8=1.0$ model. The errors of the linear bias scaling are typically a 
factor $\sim1.5-4.0$ larger than those using our fitted values of $\alpha(r)$
and $\beta(r)$.

Table~\ref{tab:delta} lists $\Delta_\up{FID}(r)$, the values of 
$\Delta\Sigma(r)$ predicted by the analytic model for the $M_r\leq-20$ and
$M_r\leq-21$ galaxy samples assuming $\Omega_m=0.3$ and $\sigma_8=0.8$. 
It also lists the $\alpha(r)$ and $\beta(r)$ functions shown in 
Figure~\ref{fig:slope}.
Equation~(\ref{eq:rel}) can be used to scale these predictions to other
values of $\Omega_m$ and $\sigma_8$, and measurements of $\Delta\Sigma(r)$
for these galaxy samples could then be used to obtain constraints in the
$\Omega_m$-$\sigma_8$ plane. 
Our prediction of $\Delta\Sigma_\up{FID}(r)$ is weakly dependent on the HOD
parametrization that we adopt when fitting $w_p(r_p)$. If we adopt the 
alternative parametrization used by \citet{idit3}, where $\alpha_\up{sat}$
is fixed to one but the $\langle N_\up{sat}\rangle_M$ cutoff is a fit 
parameter, then the $\Delta\Sigma(r)$ predictions change by less than 5\%
for all $r>0.1~\hmpc$. We have not yet explored more general parametrizations,
but we expect that the $\Delta\Sigma(r)$ predictions would be robust at this
level for all HOD models that fit the observed $w_p(r_p)$.

Galaxy-galaxy lensing measurements are often
made for flux-limited samples rather than absolute-magnitude limited samples
to increase the signal-to-noise ratio, but such measurements are difficult
to interpret quantitatively because they do not represent properties of a 
uniformly defined galaxy population. Because our predictions apply on the
non-linear scales where the measurement precision is higher, and because
results from different radii can be combined, it should be possible to
obtain precise constraints on $\sigma_8\Omega_m$ from absolute 
magnitude-limited samples, and to use different samples to check for 
consistency. Since the values of $\alpha$ and $\beta$ vary with scale, it is
possible in principle to break the degeneracy between $\Omega_m$ and 
$\sigma_8$. However, the deviations from linear scaling are not large, so 
while $\sigma_8\Omega_m$ should be well constrained, individual parameter
constraints are likely to be imprecise and sensitive to systematic 
uncertainties in the modeling.

\citet{mandel2} have recently presented SDSS measurements for narrow bins of 
luminosity and of stellar mass, which are well suited to their goal of 
constraining halo virial masses and satellite fractions. For cosmological 
parameter constraints, we think it is better to use luminosity or 
mass-threshold samples, which provide higher signal-to-noise ratio, and which
are easier to model robustly because there is no $upper$ mass cutoff on
$\langle N_\up{cen}\rangle_M$.

\section{Summary}
\label{sec:dis}
We have developed an analytic model to predict $\Delta\Sigma(r)$ for specified
cosmological and galaxy HOD parameters and tested its validity using SPH
and $N$-body simulations. We have used the analytic model to investigate the
dependence of $\Delta\Sigma(r)$ on $\sigma_8$ and $\Omega_m$ when HOD 
parameters are chosen to reproduce the observational space density and 
projected correlation function $w_p(r_p)$ of the galaxy sample being
measured. Our main findings are as follows:

1. In our SPH simulation, replacing the satellite galaxies of each halo with
randomly selected dark matter particles has a 10$-$20\% effect on 
$\xi_\up{gg}(r)$ and $\xi_\up{gm}(r)$ at scales $r\sim0.5~\hmpc$, and smaller
impact at other scales. Most of this difference arises from the differing
radial profiles of satellite galaxies and dark matter. If satellites are 
replaced in a way that preserves the radial profile but randomizes azimuthal
positions, then changes to $\xi_\up{gg}(r)$ and $\xi_\up{gm}(r)$
are $\lesssim10\%$ at all radii, and changes to $\Delta\Sigma(r)$ are $<5\%$.
Dark matter subhalos
around individual satellites orbiting in larger halos are
present, but they have negligible impact on the global $\Delta\Sigma(r)$.

2. If we randomly reassign the galaxy occupation number of each halo with 
$M<4.6\times10^{13}~\hmsun$ to another halo of nearly equal mass, then changes
to $\xi_\up{gg}(r)$, $\xi_\up{gm}(r)$, and $\Delta\Sigma(r)$ are $\lesssim2\%$
at all $r<5~\hmpc$. This result implies that any environmental dependence of
the halo occupation function $P(N|M)$ at fixed halo mass has minimal impact
on these statistics for our simulated galaxy sample, which is 
defined by a baryonic mass threshold.
For our largest scale point at $12~\hmpc$, we find an effect of
10\% on $\xi_\up{gg}(r)$, 5\% on $\xi_\up{gm}(r)$, and 2\% on 
$\Delta\Sigma(r)$, but the statistical uncertainties of our estimate are
of comparable magnitude at this scale, so larger simulation volumes are
needed to definitively establish the impact of any environmental dependence
on the large scale bias factor. Taken together, results~1 and 2 show that
the $\xi_\up{gg}(r)$ and $\Delta\Sigma(r)$ predictions of a full hydrodynamic
simulation can be reproduced to 5\% or better (usually much better)
by populating the halos of a
pure $N$-body simulation with the correct $P(N|M)$, provided that satellite
galaxy populations have the correct radial profiles.

3. Our analytic model for $\Delta\Sigma(r)$ is based on the methods introduced
by \citet{uros} and \citet{guzik1}, but it incorporates the scale-dependent
halo bias and ellipsoidal halo exclusion corrections introduced by 
\citet{zheng2} and \citet{mtl} for $\xi_\up{gg}(r)$ calculations. We have
tested the analytic model against numerical results from a grid of populated
$N$-body simulations, which span the parameter range $\sigma_8=0.6-0.95$ and
$\Omega_m=0.1-0.63$, with HOD parameters chosen to match the space density and
projected correlation function of SDSS galaxy with $M_r\leq-20$ \citep{idit3}.
The analytic model reproduces the numerical results to 5\% or better over the 
range of $0.1~\hmpc\leq r\leq20~\hmpc$. The residuals are consistent with the
statistical errors of the numerical calculations, except for the innermost bin,
where gravitational force softening in the $N$-body 
simulations artificially suppresses correlations.

4. For the $M_r\leq-20$ HOD parameters, pairs involving at least one central
galaxy dominate the galaxy-galaxy correlation function at $r\lesssim0.4~\hmpc$
and $r\gtrsim1~\hmpc$. In the range $0.4~\hmpc\lesssim r\lesssim1~\hmpc$,
satellite-satellite pairs in large halos make the dominant contribution to
$\xi_\up{gg}(r)$. In similar fashion, central galaxies dominate the 
galaxy-matter correlation function at small and large separations, while 
satellite galaxies dominate in the range 
$0.25~\hmpc\lesssim r\lesssim1.5~\hmpc$. The halos of individual satellites
make negligible contribution to $\Delta\Sigma(r)$ because they are 
usually tidally
truncated below the scales at which the satellite contribution itself is
important.

5. For samples with HOD parameters chosen to match the $M_r\leq-21$ SDSS
sample of \citet{idit3}, the ratio $\xi_\up{gm}(r)/\xi_\up{gg}(r)$ is constant
at $r\gtrsim4~\hmpc$, as predicted by the linear bias model, but it jumps
by 20$-$50\% at scales $r\sim1~\hmpc$ near the transition from the 1-halo to
2-halo clustering regime, before settling to a new, higher value at small
scales. The magnitude of the jump depends on $\sigma_8$, and it is likely to
depend on the galaxy sample as well, being stronger for more highly clustered
populations. In linear bias terms, the large scale values of 
$\xi_\up{gm}(r)/\xi_\up{gg}(r)$ correspond to a galaxy-matter cross-correlation
coefficient $r_\up{gm}\simeq0.9$, if we define the bias factor 
$b=[\xi_\up{gg}(r)/\xi_\up{mm}(r)]^{1/2}$.

6. We fit the dependence of $\Delta\Sigma(r)$ on cosmological parameters with
a scaling formula $\Delta\Sigma(r)\propto\Omega_m^\alpha\sigma_8^\beta$, 
where $\alpha$ and $\beta$ are slowly varying functions of $r$. This scaling
describes the results of our full analytic model with rms error $\lesssim3\%$
over the parameter range $\Omega_m=0.15-0.45$, $\sigma_8=0.6-1.0$. At large 
scales, $\alpha$ and $\beta$ approach the linear bias values $\alpha=\beta=1$.
However, forcing $\alpha=\beta=1$ at all scales leads to errors that are
larger by factors of $1.5-4.0$, relative to the scaling 
formula~(\ref{eq:rel}) with our fitted values of $\alpha$ and $\beta$.

Table~\ref{tab:delta} lists our predicted values of $\Delta\Sigma(r)$ for the
$M_r\leq-20$ and $M_r\leq-21$ SDSS samples, assuming our fiducial cosmological
model with $\Omega_m=0.3$ and $\sigma_8=0.8$. 
Equation~(\ref{eq:rel}) allows scaling of these results to other values of
$\sigma_8$ and $\Omega_m$, and measurements of $\Delta\Sigma(r)$ for these
samples can be combined with these predictions to obtain cosmological 
constraints, which will be tightest on a parameter combination that is 
approximately $\sigma_8\Omega_m$. Given the growth of the SDSS since the 
samples analyzed by \citet{erin} and \citet{idit3}, it is probably preferable
to extract $w_p(r_p)$ 
and $\Delta\Sigma(r)$ estimates for matched galaxy samples
from the latest data sets. A full analysis should also investigate the
effects of adding greater flexibility to the HOD parametrization itself, 
using, e.g., the 5-parameter formulation of \citet{zheng3}. We have tested the
effect of changing to a different 3-parameter description 
(see \S~\ref{sec:dep}), and we find changes of $\lesssim5\%$ in the
$\Delta\Sigma(r)$ predictions. We suspect that these predictions would remain 
similar for any choice of HOD parameters that reproduces the observed 
$w_p(r_p)$. Our SPH and $N$-body
tests indicate that the analytic model predictions should be accurate to 
5\% or better given our HOD parametrization, though assumptions about galaxy
profile concentrations have significant effect at $r<2~\hmpc$. This level
of accuracy is adequate given the statistical errors expected for current
samples, but refinement and testing on large simulations will be needed to take
full advantage of future analyses of even larger, deeper surveys.
Precise determinations of $\Omega_m$ and $\sigma_8$ can play an important
role in testing theories of dark energy and models of inflation, making 
galaxy-galaxy lensing an essential element of observational cosmology.

\acknowledgments
This work was supported by NSF grants AST-0098584 and AST-0407125,
and by NASA ATP grant NAGS-13308.
Z. Z. acknowledges the support of NASA through Hubble Fellowship grant
HF-01181.01-A awarded by the Space Telescope Science Institute, which 
is operated by the Association of Universities for Research in Astronomy,
Inc., for NASA, under contract NAS 5-26555.

\end{document}